\documentclass[11pt]{article}
\pdfoutput 1
\usepackage{graphicx}
\usepackage{amsmath}
\usepackage{amssymb}
\usepackage[colorlinks=true]{hyperref}
\usepackage{pifont}

\DeclareFontFamily{OT1}{pzc}{}
\DeclareFontShape{OT1}{pzc}{m}{it}{<-> s * [1.10] pzcmi7t}{}
\DeclareMathAlphabet{\mathpzc}{OT1}{pzc}{m}{it}

\newcommand{\vphi}{\varphi}
\newcommand{\veps}{\varepsilon}

\renewcommand{\bar}[1]{\overline{#1}}

\newcommand{\cmpq}{\widetilde{C}_{\vphi q}^{(-)33}\rule{0.0pt}{12pt}}
\newcommand{\cubr}{\widetilde{C}_{uB\,r}^{33}\rule{0.0pt}{12pt}}
\newcommand{\cubi}{\widetilde{C}_{uB\,i}^{33}\rule{0.0pt}{12pt}}
\newcommand{\cub}{\widetilde{C}_{uB}^{33}\rule{0.0pt}{12pt}}
\newcommand{\ccmpq}{C_{\vphi q}^{(-)33}\rule{0.0pt}{12pt}}
\newcommand{\ccubr}{C_{uB\,r}^{33}\rule{0.0pt}{12pt}}
\newcommand{\ccubi}{C_{uB\,i}^{33}\rule{0.0pt}{12pt}}
\newcommand{\ccub} {C_{uB}^{33}\rule{0.0pt}{12pt}}

\newcommand{\C}{\widetilde{C}}
\renewcommand{\O}{\mathcal{O}}

\title{Top quark effective couplings from top-pair tagged
  photoproduction in $pe^-$ collisions} 
\author{Antonio O.\ Bouzas and F.\ Larios\\\small Departamento de F\'{\i}sica Aplicada,
CINVESTAV-IPN \\\small Carretera Antigua a Progreso Km.\ 6, Apdo.\
Postal 73 ``Cordemex''\\\small M\'erida 97310, Yucat\'an, M\'exico}
\date{November 8, 2021}

\begin{document}

\maketitle
\begin{abstract}
  We present a detailed study, at the fast detector simulation level,
  of top-pair photoproduction in semileptonic mode at the LHeC and
  FCC-he future colliders. We work in full tree-level QED, not relying
  on the equivalent-photon approximation, taking into account the
  complete photoproduction kinematics.  This allows us to define three
  photoproduction regions based on the angular acceptance range of the
  electron tagger.  Those regions provide different degrees of
  sensitivity to top-quark effective couplings.  We focus on the
  $t\bar{t}\gamma$ dipole couplings and the left-handed vector $tbW$
  coupling for which we determine limits at both energies and in
  different photoproduction regions. We find that the LHeC and FCC-he
  will yield tight direct bounds on top dipole moments, greatly
  improving on current direct limits from hadron colliders, and direct
  limits on the $tbW$ coupling as restrictive as those expected from
  the HL-LHC.  We also consider indirect limits from $b\to s \gamma$
  branching ratio and $CP$ asymmetry, that are well known to be very
  sensitive probes of top electromagnetic dipole moments.
\end{abstract}

\newpage
\tableofcontents{}
\newpage{}

\section{Introduction}
\label{sec:intro}

Among the most important areas of research of future $pe^-$ colliders,
such as the Large Hadron-electron Collider (LHeC) and the Future
Circular Collider (FCC-he), is the study of the top quark effective
couplings to the Higgs and the electroweak bosons
\cite{LHeC2020}.  Indeed, top quark effective couplings is a
phenomenological research area of great interest
\cite{yuan1}-\cite{kozachuk}.  Top-pair and single-top production at
the LHeC are very good probes for charged-current (henceforth CC)
$tbW$ and neutral-current (NC) $ttZ$ effective couplings
\cite{sar14},\cite{mellado}.  Also, anomalous magnetic and electric
dipole moments of the top quark can be very well probed through
top-pair photoproduction in electron-proton collisions
\cite{bou13b}--\cite{billur20}.

In a previous preliminary study we obtained the dependence of the
parton-level cross section for top-pair photoproduction on the
top quark electromagnetic dipole moments, in the context of the
Standard Model Effective Field Theory (SMEFT), and
established limits on those moments.  In this paper we extend that
study by including in our Monte Carlo simulations parton showering and
hadronization, and fast detector simulation, thus making them more
realistic. This is reflected, in particular, in the study of
background processes with a variable number of jets.
Even more important, however, is the fact that in \cite{bou13b} the
cross sections for photoproduction were computed in the equivalent
photon approximation (EPA), whereas in this paper we work in full
tree-level QED, taking into account the complete kinematics of the
photoproduction process.  This extension leads to three important
improvements with respect to \cite{bou13b}. First, it leads to a
precise computation of cross sections, and of their dependence on the
photoproduction kinematic variables such as the scattered-electron
transverse momentum.  It is well known that the EPA is valid in the
logarithmic approximation and only near the reaction threshold (see
section 6.8 and appendix E of \cite{bud75}); we remove those
limitations by working in full QED.  Second, by taking account of the
complete kinematics of the process we determine the phase-space region
where there is sensitivity to the top dipole moments, and the
complementary region where there is not.  Third, we consider besides
the electromagnetic (e.m.) dipole moments, also the left-handed
vector $tbW$ coupling. We notice that in the EPA the cross-section
dependence on that coupling can occur only through the top decay
vertex, leading to very poor sensitivity. Here, the full QED
computation uncovers a phase-space region where there is significant
sensitivity to that effective coupling.

In \cite{bou13b} the top dipole moments were expressed in terms of the
Wilson coefficient $\ccub$ associated with the dimension-six effective
operator $Q_{uB}^{33}$.  Specifically, it was found in \cite{bou13b}
that either the real part of the coefficient $C^{33}_{uBr}$ at values
greater than $0.3$ or the imaginary part $C_{tBi}$ at values greater
than $0.8$ would produce a measurable $18\%$ deviation from the
Standard Model (SM) cross section.  Sometime before \cite{bou13b} was
published, we presented limits on $C^{33}_{uB}$ based on the
cross-section measurement of $t{\bar t}\gamma$ production at CDF that
were substantially weaker: $|C^{33}_{uB}|<17.0$ (either real or
imaginary part) \cite{bou13a}.  In that same report we made an
estimate of a then-future sensitivity of $t{\bar t}\gamma$ production
at the LHC at energies of $7$ or $14$TeV, and our conclusion was that
LHC should be able to set bounds of around $|C^{33}_{uB}|<3.0$
\cite{bou13a}.  Recent studies, based on ATLAS measurements have
indeed obtained limits of this size \cite{bissmannbsg}.  Also in
\cite{bou13a} we obtained limits from $BR(B\to X_s \gamma)$ and
$A_{CP}(B\to X_s \gamma)$ that were of similar size to the ones found
from LHC data.  Our analysis was based on a calculation of
$\Delta C_7 (m_W)$, the variation of the SM $C^{\rm SM}_7 (m_W)$ due
to new physics (NP) contributions from the magnetic and electric
anomalous dipole moments of the top quark found in \cite{hewett}.  A
recent study on dimension-six operators effects on $b\to s$
transitions has a new expression for $\Delta C_7 (m_W)$ that is very
different and, in particular, predicts a much higher sensitivity of
$BR(B\to X_s \gamma)$ \cite{crivellin15}.  To our knowledge there has
been no disproof of either calculation.  We shall, therefore, present
updated limits from $BR(B\to X_s \gamma)$ and the associated $CP$
asymmetry based on the two possibilities.

The structure of this paper is as follows.  In section
\ref{sec:eff.op} we discuss the effective Lagrangian framework of our
study. We summarize there, also, the results of several recent global
analyses of top quark effective couplings in that framework. In
section \ref{sec:php.sm} we carry out a Feynman-diagram analysis of
the top-pair photoproduction process in $pe^-$ collisions in the
SM. In section \ref{sec:php.mc} we describe in detail the Monte Carlo
simulations of the photoproduction process and its main irreducible
background and obtain their cross sections in the SM.  In section
\ref{sec:sm.bck} we enumerate several additional SM background
processes and assess their relative importance. In section
\ref{sec:res.coupl} we present the limits to the top e.m.\ dipole
moments and left-handed vector $tbW$ coupling at the LHeC and FCC-he
energies in different photoproduction kinematical regions. In section
\ref{sec:final} we give a summary of the paper and our final remarks.
Several appendixes provide additional technical details on some issues
discussed in the main text.  Appendix \ref{sec:neutr.reco} deals with
our approach to neutrino momentum reconstruction.  In appendix
\ref{sec:b.sa} we give the details on the calculation of
effective-coupling limits from $BR(B\to X_s \gamma)$ and its
associated asymmetry. In appendix \ref{sec:app.bound} we restate some
of the main results from section \ref{sec:res.coupl} in a different
convention to facilitate comparison with other studies.

\section{Effective SM Lagrangian}
\label{sec:eff.op}

The framework we use in this paper is the SM effective Lagrangian
with operators up to dimension six. In this context, the Lagrangian
for top-pair photoproduction is of the form 
\begin{equation}
  \label{eq:lag}
  \mathcal{L} = \mathcal{L}_\mathrm{SM} + \frac{1}{\Lambda^2}\sum_{\mathcal{O}}
  (\widehat{C}_\mathcal{O} \mathcal{O}+\mathrm{h.c.})+\cdots, 
\end{equation}
where $\mathcal{O}$ denotes dimension-six effective operators, $\Lambda$
is the new-physics scale, and the ellipsis refers to
higher-dimensional operators. It is understood in (\ref{eq:lag}) that
the addition of the Hermitian conjugate, denoted $+\mathrm{h.c.}$ in
the equation, is applicable only to non-Hermitian operators.
Throughout this paper we use the dimension-six effective operators from
the operator basis given in \cite{grz10}.  In particular, we use the
same sign convention for covariant derivatives as in
\cite{bou13b,grz10}, namely, $D_\mu=\partial_\mu + i e A_\mu$ for the
electromagnetic coupling.  However, we adopt the operator
normalization defined in \cite{zha14} (see also \cite{manohar1309}),
where a factor $y_t$ is attached to an operator for each Higgs field
it contains, and a factor $g$ ($g'$) for each $W_{\mu\nu}$
($B_{\mu\nu}$) field-strength tensor.  The Wilson coefficients in
(\ref{eq:lag}) are denoted $\widehat{C}$, since we will denote $C$ the
coefficients associated with the original operator basis \cite{grz10}.
In fact, it will be convenient in what follows to express our results
in terms of the modified dimensionless couplings
\begin{equation}
  \label{eq:coupl}
  \C_\mathcal{O} = \widehat{C}_\mathcal{O} \frac{v^2}{\Lambda^2},
\end{equation}
where $v$ is the Higgs-field vacuum expectation value.  At tree level
the coupling constants $\C_\mathcal{O}$ are independent of the scale
$\Lambda$. We denote complex couplings as
$\C_{\mathcal{O}}=\C_{\mathcal{O}\,r}+i \C_{\mathcal{O}\,i}$.

There are seven operators in the basis \cite{grz10} that couple
electroweak bosons and third family quarks.  One of them is the focus
of our study: $Q^{33}_{uB}$.  Another one, $Q^{33}_{\varphi u}$,
generates only a right-handed $ttZ$ coupling so that it will not
contribute to the photoproduction process.  Other four of them
generate anomalous $tbW$ couplings: $Q^{33}_{\varphi ud}$,
$Q^{33}_{uW}$, $Q^{33}_{dW}$ and $Q^{(3)33}_{\varphi q}$. The first
three receive strong limits from top-decay $W$-helicity fractions
\cite{atl12,cms13,cms14} and, since the expected sensitivity of
top-pair photoproduction to those couplings is low, we will not
consider them further here. Unlike those three,
$Q^{(3)33}_{\varphi q}$ only gets loose bounds from $W$-helicity
fractions measurements. In a previous study we have obtained the LHeC
sensitivity to the $tbW$ couplings from single-top production and
indeed we have found that the HL-LHC would give stronger (individual)
constraints on the tensorial $Q^{33}_{uW}$ and $Q^{33}_{dW}$ operators
\cite{sar14}.  On the other hand, the LHeC would have better
sensitivity than the HL-LHC for $Q^{(3)33}_{\varphi q}$ \cite{sar14}
(and to a lesser degree, also for $Q^{33}_{\varphi ud}$).
To date, the strongest direct bounds to $Q^{(3)33}_{\varphi q}$ come
from single-top production cross sections at the LHC \cite{cms17a}.
Besides $Q^{33}_{uB}$, which is the focus of our study, we will also
include the left-handed $tbW$ operator $Q^{(3)33}_{\varphi q}$ in
our analysis because, as we show below, $t\bar t$ photoproduction at
the LHeC and FCC-he can also be a competitive probe for this coupling.
However, a redefinition of this operator that is also commonly used in
the literature \cite{zhangmemory} will be necessary.
$Q^{(3)33}_{\varphi q}$ and $Q^{(1)33}_{\varphi q}$ both generate
$ttZ$ and $bbZ$ couplings. As the latter receives very strong
constraints from bottom-quark production at the $Z$-pole (LEP and
SLAC), we will have to combine the operators $Q^{(3)33}_{\vphi q}$ and
$Q^{(1)33}_{\vphi q}$ so as to eliminate the $bbZ$ neutral current
term \cite{yuan2,yuan3}.  Thus, those operators will not be considered by themselves but
rather in a particular linear combination.

In this paper we will, therefore, set up our analysis by focusing on
just two operators: $\O_{uB}^{33}$ and $\O^{(-)33}_{\vphi q}$. \footnote{
  It should noted that, eventually, a complete global analysis should be made
  of the combined HL-LHC/LHeC sensitivity to all the operators
  mentioned here.}
Expanding these operators in physical fields yields, with the
conventions discussed above:
\begin{equation}
  \begin{aligned}
\O^{33}_{uB} &= y_t g' Q^{33}_{uB}= \sqrt{2} y_t e (v+h) (\partial_\mu
A_\nu - \tan\theta_W \partial_\mu Z_\nu)\; \bar{t}_{L}\sigma^{\mu\nu}
t_{R}~,\\    
\O_{\vphi q}^{(-)33} &= \O_{\vphi q}^{(3)33}-\O_{\vphi q}^{(1)33} =
-y_t^2 Q_{\vphi q}^{(-)33} \\
&=-y_t^2\frac{g}{\sqrt{2}} (v+h)^2 \left( W^+_\mu\, \bar{t}_{L}
  \gamma^\mu b_L +
  W^-_\mu\,\bar{b}_{L} \gamma^\mu t_L \right) - y_t^2\frac{g}{c_W} (v+h)^2 Z_\mu\, \bar{t}_{L}\gamma^\mu t_{L} \, ,\\
  \end{aligned}
  \label{eq:operators}
\end{equation} 
where $Q^{33}_{uB}$, $Q_{\vphi q}^{(-)33}$ are the operators defined in
\cite{grz10}.  Notice that both operators $\O^{33}_{uB}$ and
$\O_{\vphi q}^{(-)33}$ are $O(g^1)$ with respect to the weak coupling
constant, which makes the definitions (\ref{eq:operators}) consistent
from the point of view of perturbation theory.  We stress here the
definition
$\O_{\vphi q}^{(-)33} = \O_{\vphi q}^{(3)33}-\O_{\vphi q}^{(1)33}$ we
use, since sometimes in the literature the opposite sign is used.
The effective Lagrangian used throughout this paper results from
substituting (\ref{eq:operators}) and (\ref{eq:coupl}) in the
Lagrangian (\ref{eq:lag}).  It is convenient to record here the
relation between the Wilson coefficients in the form (\ref{eq:coupl})
and those associated with the original basis \cite{grz10} (see also
\cite{newcite}).  From the relation
\begin{equation*}
  \mathcal{L} = \mathcal{L}_\mathrm{SM} + \frac{1}{\Lambda^2}\sum_{\mathcal{Q}}
  (C_Q Q+\mathrm{h.c.})+\cdots, 
\end{equation*}
analogous to (\ref{eq:lag}), we obtain
\begin{equation}
  \label{eq:coupl2}
  C_{uB}^{33} = \frac{\Lambda^2}{v^2} y_t g'\cub = 5.906 \cub~,
  \qquad
  C_{\varphi q}^{(-)33} = -\frac{\Lambda^2}{v^2} y_t^2 \cmpq = -16.495
  \cmpq~. 
\end{equation}
The numerical values in this equation arise from the parameters
$\Lambda=1$ TeV, $v=246$ GeV, $g'=0.358$, $g=0.648$. Furthermore, for
all practical purposes we set $y_t=1=V_{tb}$. We point out here that
the couplings $\cmpq$, $\cub$ are both of $O(g^0)$ in the perturbative
expansion and, therefore, from (\ref{eq:coupl2})
$C_{\varphi q}^{(-)33}$ is $O(g^0)$ but $C_{uB}^{33}$ is $O(g^1)$.

It is common practice in the literature to write the anomalous 
interactions in terms of form factors.  We adopt here the definition of
top electromagnetic dipole moments given in eq.\ (2) of \cite{bou13b},
and the CC vertex form factors from eq.\ (7.1) of \cite{cms17a},
\begin{equation}
  \label{eq:form.fact}
  \begin{aligned}
  \mathcal{L}_\mathrm{anom} &= \mathcal{L}_\mathrm{anom,em} + \mathcal{L}_\mathrm{anom,CC}~,\\
  \mathcal{L}_\mathrm{anom,em} &= \frac{e}{4m_t}\bar{t}\,\sigma^{\mu\nu}(\kappa
  + i \widetilde{\kappa}\gamma_5)\,t\; F_{\mu\nu}~,\\
   \mathcal{L}_\mathrm{anom,CC} &=  \frac{g}{\sqrt{2}} f_V^L \left(
    W^+_\mu (\bar{t}_{L}\gamma^\mu b_{L}) + W^-_\mu
    (\bar{b}_{L}\gamma^\mu t_{L}) \right)~,\quad f_V^L = V_{tb} + \delta f_V^L~.
\end{aligned}  
\end{equation}
Direct comparison of (\ref{eq:form.fact}) with (\ref{eq:lag}),
(\ref{eq:coupl}), (\ref{eq:operators}), yields the tree-level
relations,
\begin{equation}
  \label{eq:relations}
    \kappa = 2 y_t^2 \widetilde{C}_{uB\,r}^{33},
  \qquad
  \widetilde{\kappa} = 2 y_t^2 \widetilde{C}_{uB\,i}^{33},
  \qquad
    \delta f_V^L = y_t^2 \widetilde{C}_{\vphi q}^{(-)33}~.
\end{equation}
The sign chosen in the last equality in (\ref{eq:relations}) deserves
clarification.  Because the signs in the corresponding operator
definitions (\ref{eq:operators}) and (\ref{eq:form.fact}) are
opposite, a relative ``$-$'' sign could be expected in the relation
between $\delta f_V^L$ and $\widetilde{C}_{\vphi q}^{(-)33}$.
However, we define the SM CC Lagrangian in this paper with the same
sign convention as in (\ref{eq:operators}), and we assume that the
analogous convention is made in \cite{cms17a}, with the result that
the interference term between the effective CC interaction and the SM
one has the same sign (``$+$'') in both cases.  Thus,
(\ref{eq:relations}) gives the correct relationship between
$\delta f_V^L$ and $\widetilde{C}_{\vphi q}^{(-)33}$.
We also notice here, that the particularly simple relations
(\ref{eq:relations}) are a consequence of
eq.\ (\ref{eq:coupl}) and
the operator normalization discussed above in the paragraph
under eq.\ (\ref{eq:lag}).

\subsection{Overview of global analyses on top quark
  effective couplings.}
\label{sec:overview}

In the last few years several research groups have reported extensive
analyses of limits on certain sets of SMEFT dimension-six operators
based, in turn, on certain sets of experimental observables.  It is
well known that no one experimental observable is only related to only
one effective operator in isolation.  Rather, for any observable there
are usually four, five or many more independent operators that
contribute.  So the goal is not necessarily to prove that a certain
observable is the best candidate to test one coupling in particular,
but that this observable can contribute significantly to the pool of
measurements that will be used in future ever more extensive analyses.

In a recent study  constraints on the three top dipole operators
$Q_{tB}$, $Q_{tW}$ and $Q_{tG}$ are obtained by using as experimental
inputs the branching ratio $BR ({\bar B} \to X_s \gamma)$ and two
fiducial cross-section results of $t{\bar t}\gamma$ production
by ATLAS \cite{bissmannbsg}.
It is found that, despite the large difference in sensitivities
between $t{\bar t}\gamma$ and ${\bar B} \to X_s \gamma$ the inclusion
of the less sensitive $t{\bar t}\gamma$ input still improves
significantly the combined marginalized constraints.
As far as experimental inputs the 
$BR({\bar B} \to X_s \gamma)$ stands out as probably the one
observable that is most sensitive to the magnetic dipole
operator $Q_{tB}$.   However, we must also be aware of the
fact that this is an indirect observable that is actually
sensitive to most of the dimension-six operators with quark
fields of the Warsaw basis \cite{crivellin15}.  And even so,
$Q_{tB}$ is among those operators whose contribution to the
Wilson coefficients of the effective Hamiltonian is not at
the tree but at the one-loop level \cite{crivellin15}.
It is well known that direct observables, like
production cross sections, their kinematic distributions and
other related observables measured by colliders like the
Tevatron and the LHC will always play the essential part of
effective coupling studies.

Let us consider the reports of the last two years on global fits to
gauge and Higgs boson combined with top and bottom quarks in the
literature.  In table~\ref{tab:global} we show some of the limits
reported by five collaborations: SMEFiT \cite{zhangmemory}, TopFitter
\cite{brown}, Fitmaker \cite{ellis}, HEPfit \cite{miralles} and
EFTfitter \cite{bissmann}. Out of these five groups only the last one
uses $B$ meson observables, while the rest rely mostly on ATLAS and
CMS data on single top, $t\bar t$, $W$-helicity in top decay, and
similar measurements of top quark processes.  It is interesting to
observe the very diverse scenarios they depict for individual
constraints, as for instance Fitmaker and HEPfit obtain very stringent
bounds on $C^{(3)}_{\vphi Q}$ as compared to SMEFiT and TopFitter, but
then the exact opposite is true for $C_{tW}$ again among these four
collaborations.  On the other hand, once the global fit is performed
and marginalized limits are obtained we do see a more uniform
scenario, as is shown in the lower part of table~\ref{tab:global}.
Since we are interested in the electromagnetic dipole coupling, we
added the tag $t{\bar t}\gamma$ to the last three groups that used
this LHC process in their fits, and then see if they would be the ones
with better constraints on $C_{tB}$.  Apparently, this is not quite
the case, TopFitter and Fitmaker end up with similar limits despite
one not relying on $t{\bar t}\gamma$.  Of course, the best bounds
(which are not even individual) appear in the last column; but we know
that their strong bound really comes from $\bar B \to X_s \gamma$.  We
point out that the limits shown in Table 1 have been obtained using
linear (or interference) contributions as well as quadratic (or purely
anomalous) contributions to the chosen observables.  These bounds are
usually somewhat more stringent than the ones obtained with only
linear terms.  The exception is in the middle column, where the limits
provided by Fitmaker \cite{ellis} are only with linear terms.

\begin{table} \centering{}
  \begin{tabular}{|c|c|c|c|c|c|}
    \cline{1-6} $C$ & SMEFiT \cite{zhangmemory} &
 TopFitter \cite{brown} & Fitmaker \cite{ellis} &
 HEPfit \cite{miralles} & EFTfitter \cite{bissmann} \\
 & & &$t{\bar t}\gamma$&$t{\bar t}\gamma$&$t{\bar t}\gamma$,B\\
 \hline
 & \multicolumn{4}{c}{\mbox{individual 90\% - 95\% C.L.}}& \\
 \hline
$C^{(1)}_{\vphi Q}$ &$--$&$-3.0\;,\;+0.8$&
    $-0.03\;,\;+0.05$&$--$&$--$\\
$C^{(3)}_{\vphi Q}$ &$-0.38\;,\;+0.34$&$-0.3\;,\;+0.9$&
    $-0.03\;,\;+0.05$&$-0.02\;,\;+0.04$&$--$\\
$C^{(-)}_{\vphi Q}$ &$-1.1\;,\;+1.6$&$--$&
    $--$&$-0.04\;,\;+0.08$&$--$\\
$C^{}_{\vphi t}$ &$-3.0\;,\;+2.2$&$-1.0\;,\;+4.5$&
    $-1.2\;,\;+2.9$&$-8.6\;,\;+1.5$&$--$\\
$C^{}_{\vphi tb}$ &$--$&$-0.3\;,\;+0.3$&
    $--$&$-6.6\;,\;+6.7$&$--$\\
$C^{}_{tW}$ &$-0.08\;,\;+0.03$&$-0.09\;,\;+0.09$&
    $-0.12\;,\;+0.51$&$-0.28\;,\;+0.32$&$--$\\
$C^{}_{bW}$ &$--$&$-0.05\;,\;+0.05$&
    $--$&$-0.47\;,\;+0.47$&$--$\\
$C^{}_{tZ}$ &$-0.04\;,\;+0.09$&$--$&
    $--$&$-0.39\;,\;+0.57$&$--$\\
$C^{}_{tB}$ &$--$&$-4.2\;,\;+4.5$&
    $-4.5\;,\;+1.2$&$--$&$--$\\
 \hline
 & \multicolumn{4}{c}{\mbox{marginalized 90\% - 95\% C.L.}}& \\
 \hline
$C^{(1)}_{\vphi Q}$ &$--$&$--$&$-0.59\;,\;+0.58$
 &$--$&$\pm 0.43$\\
$C^{(3)}_{\vphi Q}$ &$-0.62 \;,\;+0.48$&$--$&
  $-0.67\;,\;+0.46$&$-1.29\;,\;+0.81$&$\pm 0.40$\\
$C^{(-)}_{\vphi Q}$ &$-2.25\;,\;+2.86$&$--$&
    $--$&$-2.42\;,\;+2.29$&$--$\\
$C^{}_{\vphi t}$ &$-13.36\;,\;+3.96$&$--$&
    $+2\;,\;+11$&$-10.58\;,\;+1.12$&$\pm 13.2$\\
$C^{}_{\vphi tb}$ &$--$&$--$&
    $--$&$-7.6\;,\;+7.6$&$--$\\
$C^{}_{tW}$ &$-0.24\;,\;+0.09$&$--$&
$-0.09\;,\;+0.55$&$-0.19\;,\;+0.50$&$\pm 0.50$\\
$C^{}_{bW}$ &$--$&$--$&
 $--$&$-0.98\;,\;+0.94$&$--$\\
$C^{}_{tZ}$ &$-1.13\;,\;+0.86$&$--$&
    $--$&$-0.37\;,\;+0.88$&$--$\\
$C^{}_{tB}$ &$--$&$--$&
    $-5.2\;,\;+2.5$&$--$&$\pm 0.58$\\
    \hline
    \end{tabular}
    \caption{
Limits on the effective couplings at 90-95\% C.L.
from global fits reported in the last two years.
Coefficients as defined in \cite{grz10}, and
$C^{(1,3)}_{\vphi Q}= C^{(1,3)33}_{\vphi q}$,
$C^{(-)}_{\vphi Q}= C^{(1)33}_{\vphi q} - C^{(3)33}_{\vphi q}$,
$C_{tB} = C^{33}_{uB}$, $C_{tW} = C^{33}_{uW}$,
$C_{tB} = C^{33}_{uB}$, $C_{tZ} = c_w C^{33}_{uW} - s_w C^{33}_{uB}$.}
  \label{tab:global}
  \end{table}

Concerning future projections for the limits of $C^{33}_{uB}$,
or $\kappa$, $e^+ e^-$ colliders like CLIC and ILC offer the highest
sensitivity with potential bounds of order $|\kappa| \leq 0.003$
($|C_{uB}^{33}| \leq 0.009$) and $|\kappa| \leq 0.002$ ($|C_{uB}^{33}|
\leq 0.006$), respectively \cite{bernreuther17}.  For the HL-LHC
$tt\gamma$ a preliminary study in \cite{baur04} (see also
\cite{baur05}) obtained $|\kappa| \leq 0.12$ ($|C_{uB}^{33}| \leq 0.36$)
for $3 {\rm ab}^{-1}$ luminosity.  Recently, a more realistic analysis
by \cite{ATLAS2018} obtained a similar sensitivity for
$3 {\rm ab}^{-1}$ luminosity, with $-0.5 \leq C_{uB}^{33} \leq 0.3$.  These
HL-LHC potential limits are similar to the bounds obtained here for
LHeC, and also of similar size as the current $68\%$ C.L. individual
bound from ${\bar B} \to X_s \gamma$ as shown in Appendix
\ref{sec:b.sa}.

\section{Top-pair photoproduction in the SM: diagrammatic
analysis}
\label{sec:php.sm}

\begin{figure}
  \centering{}
  \includegraphics[scale=0.75]{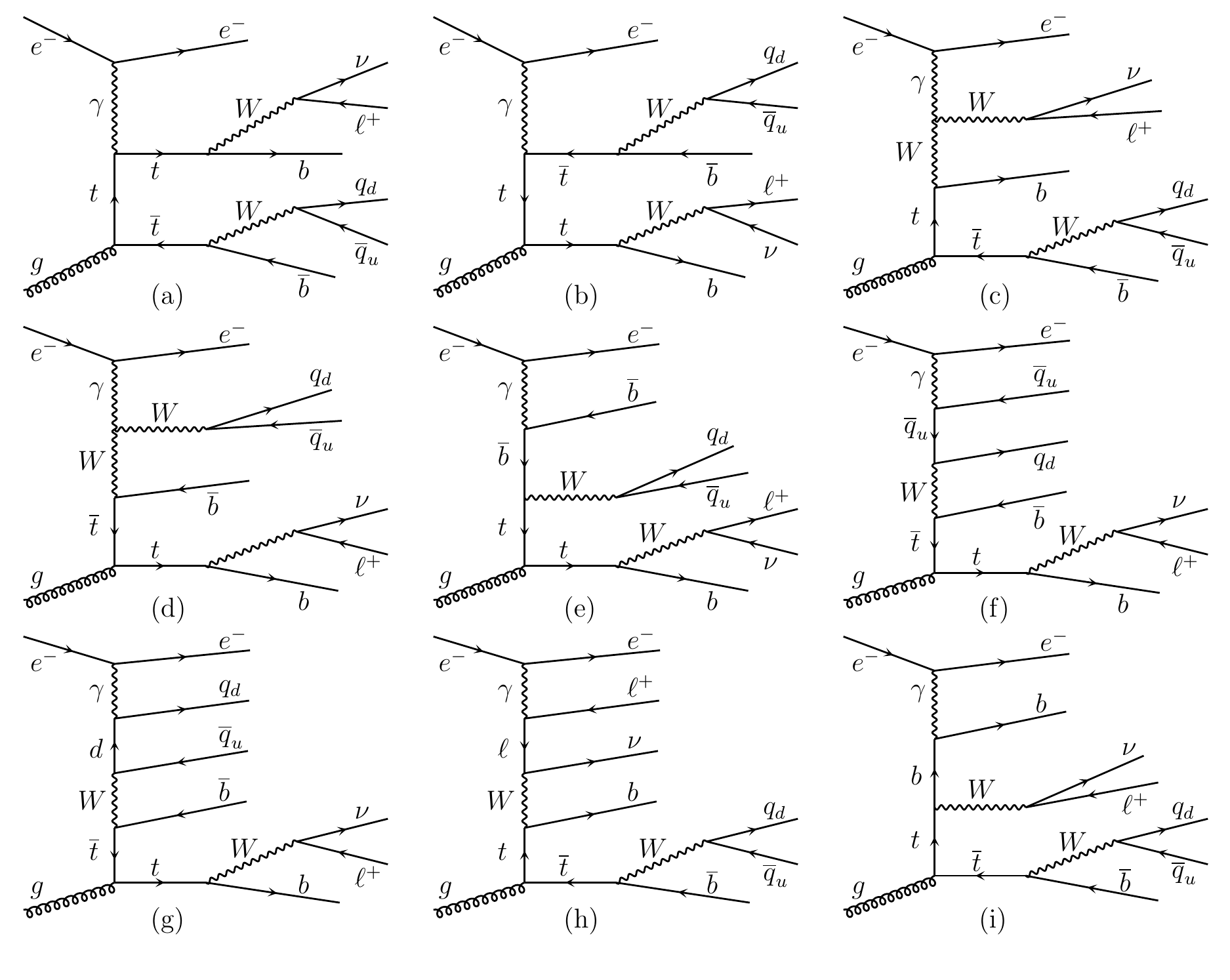}
  \caption{Unitary-gauge Feynman diagrams for the photoproduction of a
    top pair in semileptonic mode, see eq.\ (\ref{eq:ttx.proc}).
    \mbox{All~diagrams} for the final state
    $e^-\, b\ell^+\nu_\ell\, \bar{b}\bar{q}_uq_d$ are shown.  Diagrams
    (c)--(i) are necessary to preserve electromagnetic gauge
    invariance when $t$, $W$ are off shell.}
  \label{fig:feyn.1}
\end{figure}
We are interested in this paper in top-pair photoproduction in $pe^-$
collisions in the semileptonic decay channel which, at parton level,
leads to the seven-fermion final states,
\begin{equation}
  \label{eq:ttx.proc}
    g\; e^- \rightarrow e^- t\bar{t}\rightarrow
    e^-\,b\ell^+ \nu_\ell\, \overline{b}\overline{q}_u q_d +
    e^-\, bq_u \overline{q}_d\,\overline{b}\ell^- \overline{\nu}_\ell,
    \quad \mathrm{with}\; q_u=u,c, \; q_d=d,s,\; \ell=e,\mu.
\end{equation}
This equation defines our signal process.  The set of Feynman diagrams
for this process in the photoproduction region in unitary gauge in the
SM with Cabibbo mixing is shown in figure \ref{fig:feyn.1}. For each
possible final-state lepton ($\ell=e^\pm$, $\mu^\pm$) in
(\ref{eq:ttx.proc}) there are four possible quark-flavor combinations,
and for each possible lepton and quark flavor combination there are
nine diagrams in figure \ref{fig:feyn.1}, except for $\ell=e^-$ for
which the number of diagrams doubles, leading to a total of 180
Feynman diagrams for this process. We consider the top-pair
photoproduction process defined by the diagrams in figure
\ref{fig:feyn.1} our signal process, and denote its cross section by
$\sigma_\mathrm{sgnl}$.

We can divide the set of diagrams in figure \ref{fig:feyn.1} into two
subsets: the first subset includes diagrams (a), (b), containing three
internal top lines, and the second one comprises the remaining
diagrams, (c)--(i), containing two internal top lines. The second
subset is necessary to preserve electromagnetic gauge invariance in
the phase-space regions where $t$ or $W$ lines are off shell. There
is, in fact, strong destructive interference between the two subsets
in the photoproduction region, such that the total cross section
$\sigma_\mathrm{sgnl}$ computed with all the diagrams is smaller than
the cross sections obtained from (a), (b) or (c)--(i) separately, by a
factor 10--25 depending on the cuts used in the computation.

\begin{figure}
  \centering{}
  \includegraphics[scale=0.72]{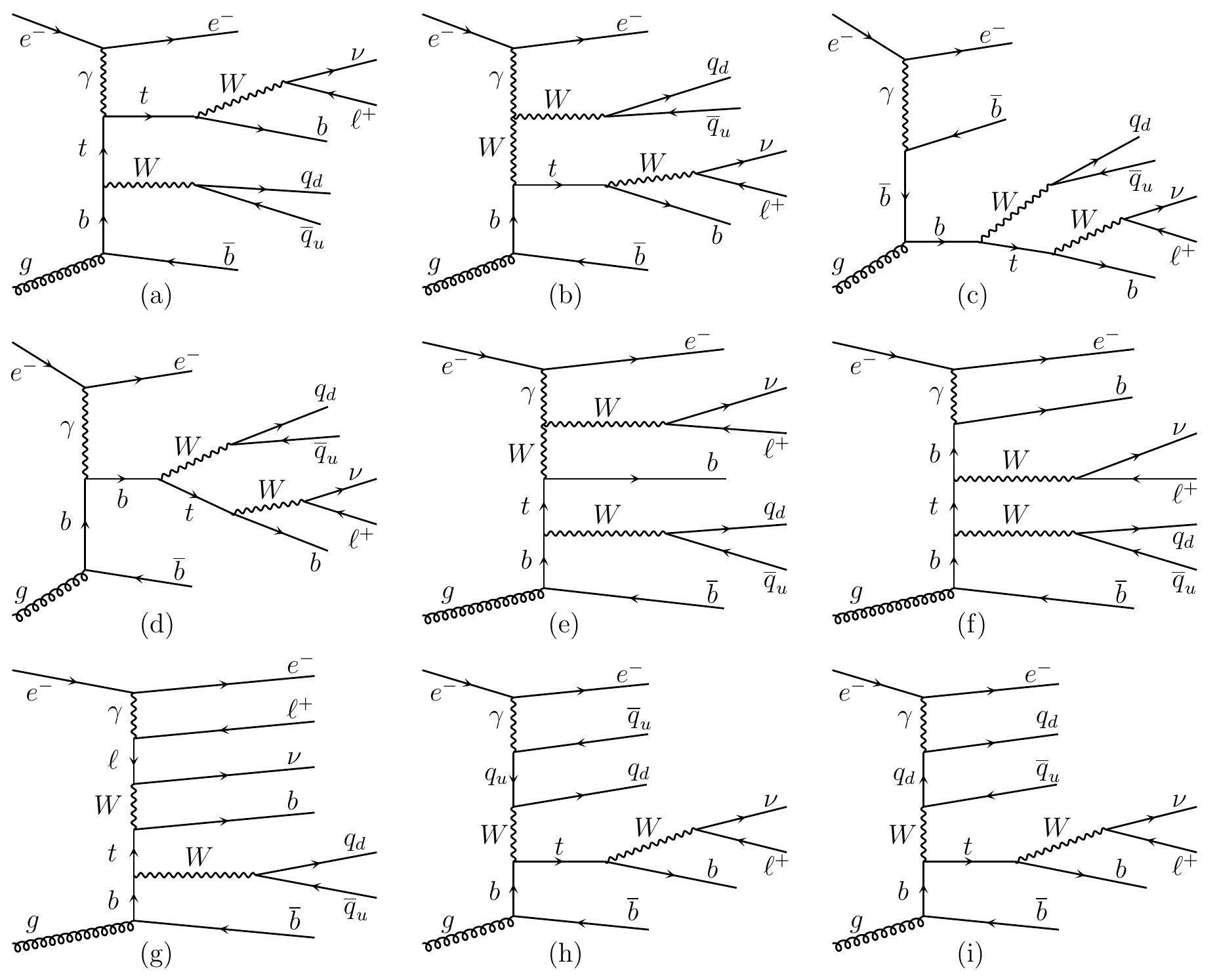}
  \caption{Unitary-gauge Feynman diagrams for the associate
    photoproduction and decay of $tbW$ in semileptonic mode with
    leptonic top decay, see eq.\ (\ref{eq:tbw.proc}).
    \mbox{All~diagrams} for the final state
    $e^-\, b\ell^+\nu_\ell\, \bar{b}\bar{q}_uq_d$ corresponding to
    leptonic top decay are shown. Diagrams (e)--(i) are necessary to
    preserve electromagnetic gauge invariance when $t$, $W$ are
    off shell.}
  \label{fig:feyn.2}
\end{figure}
We must consider also other processes with the same final state as
(\ref{eq:ttx.proc}), which constitute irreducible backgrounds.
Particularly important is the associate $tbW$ photoproduction, in
which $bW$ does not originate in a top decay.  For this process,
in semileptonic mode, we have to distinguish the cases of leptonic and
hadronic top decays,
\begin{equation}
  \label{eq:tbw.proc}
  \begin{aligned}
  ge^- &\rightarrow e^- t\bar{b}W^- + e^- \bar{t}bW^+\rightarrow
  e^-\,b\ell^+\nu_\ell\,\bar{b}\bar{q}_uq_d
  +e^-\,\bar{b}\ell^-\bar{\nu}_\ell\,bq_u\bar{q}_d,\\
  ge^- &\rightarrow e^- t\bar{b}W^- + e^- \bar{t}bW^+\rightarrow
  e^-\,bq_u\bar{q}_d\,\bar{b}\ell^-\bar{\nu}_\ell
  +e^-\,\bar{b}\bar{q}_uq_d\,b\ell^+\nu_\ell,  
  \end{aligned}
\end{equation}
with $q_u$, $q_d$, $\ell$ as in (\ref{eq:ttx.proc}).  These two sets
of processes lead to the same final states, and to two different but
completely analogous sets of Feynman diagrams.  In figure
\ref{fig:feyn.2} we show the corresponding diagrams for $t\bar{b}W^-$
photoproduction with leptonic $t$ decay.  Taking into account the
possible lepton and quark flavor channels, the duplication of diagrams
for $\ell=e^-$, and the two types of processes in (\ref{eq:tbw.proc}),
we are led to a total of 360 Feynman diagrams for
(\ref{eq:tbw.proc}). There is strong destructive interference in the
photoproduction region, similar to that described above for the signal
process, between the subset of diagrams from figure \ref{fig:feyn.2}
formed by diagrams (a)--(d) and that formed by (e)--(i). The latter
is needed to preserve electromagnetic gauge invariance when $t$ or $W$
are off shell.  This process, as discussed in detail below, is the
main irreducible background to the signal process (\ref{eq:ttx.proc})
and we denote its cross section by $\sigma_{tbW}$.

\begin{figure}
  \centering{}
  \includegraphics[scale=0.72]{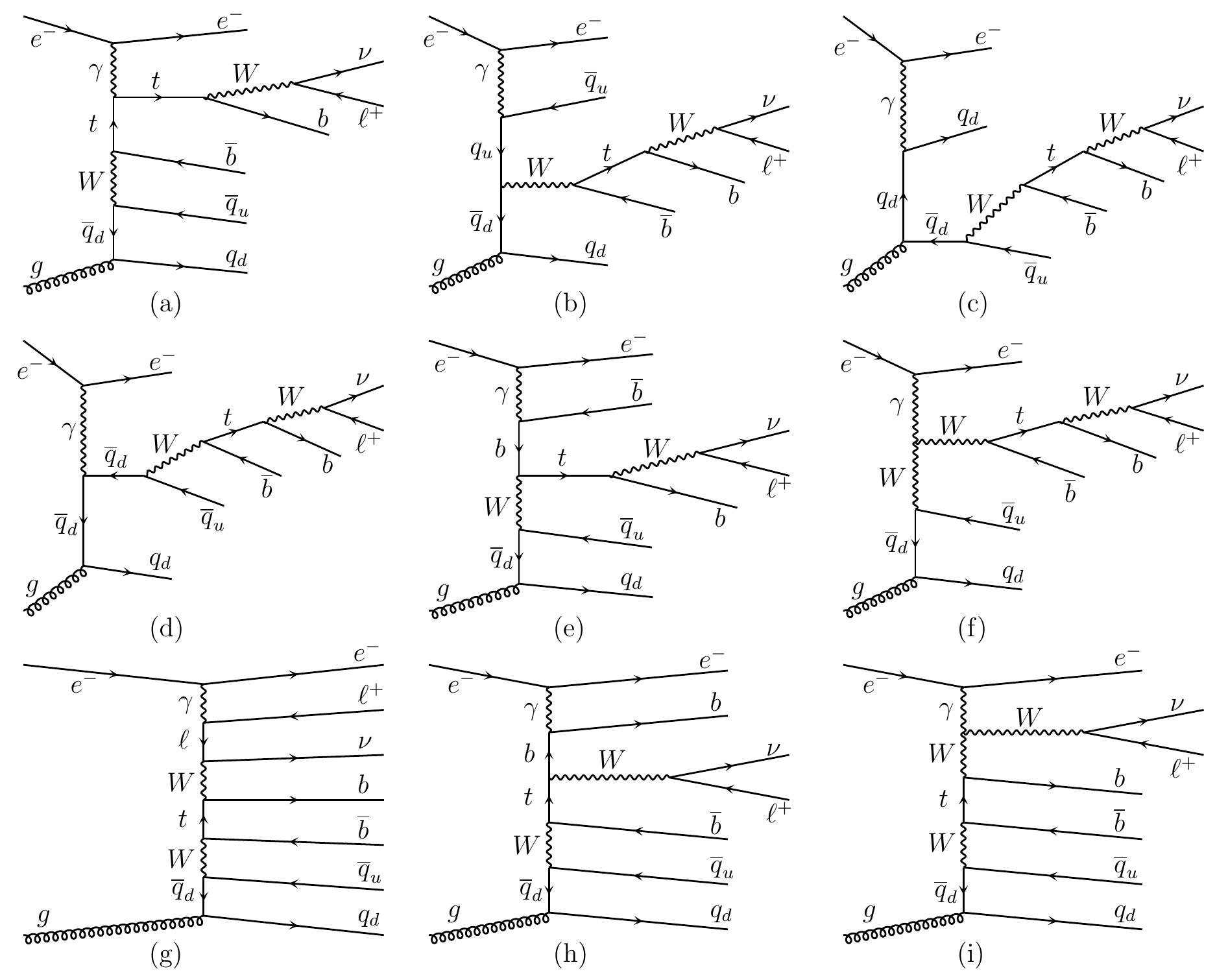}
  \caption{Unitary-gauge Feynman diagrams for the associate
    photoproduction and decay of $tbqq$ in leptonic mode, see eq.\
    (\ref{eq:tbqq.proc}).  \mbox{All~diagrams} for the final state
    $e^-\, b\ell^+\nu_\ell\, \bar{b}\bar{q}_uq_d$ are shown. Diagrams
    (g)--(i) are necessary to preserve electromagnetic gauge
    invariance when $t$, $W$ are off shell.}
  \label{fig:feyn.3}
\end{figure}
Another process with the same final state as (\ref{eq:ttx.proc}) that
we must take into account is $tbqq$ photoproduction with leptonic top
decay, in which the $qq$ does not arise from a $W$ decay,
\begin{equation}
  \label{eq:tbqq.proc}
  ge^-\rightarrow e^- t\bar{b}\bar{q}_uq_d
  + e^- \bar{t}bq_u\bar{q}_d
  \rightarrow
    e^-\,b\ell^+\nu_\ell\,\bar{b}\bar{q}_uq_d +
    e^-\,\bar{b}\ell^-\bar{\nu}_\ell\,bq_u\bar{q}_d,
\end{equation}
with $q_u$, $q_d$, $\ell$ as in (\ref{eq:ttx.proc}). In figure
\ref{fig:feyn.3} we show the Feynman diagrams for the process
(\ref{eq:tbqq.proc}) in its $t\bar{b}\bar{q}_uq_d$ form, containing a
$gq_d\bar{q}_d$ vertex. From those diagrams, by exchanging
$q_d\leftrightarrow \bar{q}_u$, $\bar{q}_d\leftrightarrow q_u$,
another set of valid ones with a $gq_u\bar{q}_u$ vertex can be
obtained, which is not shown for brevity. Taking account of the quark-
and lepton- flavor multiplicity as above, and the diagrams with
$gq_u\bar{q}_u$ vertex not shown, results in a total of 360 diagrams
for process (\ref{eq:tbqq.proc}).  As discussed below, $tbqq$
constitutes a small background to $t\bar{t}$ photoproduction that can
be neglected.

We stress here that the diagrams for processes (\ref{eq:ttx.proc}),
(\ref{eq:tbw.proc}), (\ref{eq:tbqq.proc}) exhaust all possible Feynman
diagrams (900 in total) with the same initial and final states as
those processes and with at least one internal top line, in the
photoproduction region, in unitary gauge, in the SM with Cabibbo
mixing.  Furthermore, the interference between those processes is
small, at the level of a few percent, so that it makes sense to
consider $t\bar{t}$ photoproduction the signal process with $tbW$ and
$tbqq$ as backgrounds.

\section{Top-pair photoproduction in the SM: Monte Carlo simulations}
\label{sec:php.mc}

We compute the tree-level cross section for top-pair photoproduction
and its backgrounds with the matrix-element Monte Carlo generator
MadGraph5\_aMC@NLO (henceforth MG5) version 2.6.3 \cite{alw14}.  We
use the parton distribution function (PDF) CTEQ6l as implemented in
MG5.  The cuts we apply at the parton level in MG5 are similar to
those described below in connection with event selection, but
substantially looser, in order to adequately populate phase space
without inappropriately restricting it.  As should be clear from the
discussion in section \ref{sec:php.sm}, all resonant and non-resonant
Feynman diagrams for this process are taken into account, as well as
all off-shell and interference effects.  In particular, the
small-width approximation is not used in our simulations. We run
Pythia version 6.428 \cite{sjo06} with MG5 events as initial data,
with default parameters, for QCD/QED showering, hadronization and
resonance decay. In the parton-level MG5 simulation we work in the
four light flavors scheme, retaining the masses of the first two
particle generations to keep Pythia's event rejection rate at 0. We
neglect, however, the Higgs boson couplings to the first two
generations for numerical efficiency, since the light-mass effects are
negligibly small.

We run Delphes version 3.4.2 \cite{fav14} on Pythia events for fast
detector simulation.  Jet clustering is carried out in Delphes by
means of FastJet 3.3.2 \cite{cac12}.  We carry out the analysis of
Delphes events with Root version 6.22 \cite{root}. For the LHeC and
FCC-he detector simulations we use the configuration files developed
by the experimental collaborations and distributed with Delphes as
delphes\_card\_LHeC.tcl and delphes\_card\_FCCeh.tcl, respectively.
In both cases we use default parameters, with the following
exceptions.
For jet clustering we use the anti-$k_t$ algorithm with default radius
parameter 0.4, but we set \texttt{JetPTMin} to 10 GeV since we will
have to introduce a cut on jet $p_T$ with $p_{T\,\mathrm{min}}\geq 10$
GeV to control irreducible backgrounds.  Furthermore, we find that at
the LHeC the cross section for top-pair photoproduction is somewhat
small, and that reducible backgrounds are much less important than
irreducible ones.  We are thus led to choose a $b$-tagging algorithm
with higher efficiency and lower purity than the default.  We
therefore change the settings in the LHeC configuration file
from the default with $b$-tagging efficiency $\eta_b=0.75$,
and $c$- and light-jet mistagging probabilities $p_c=0.05$ and
$p_j=0.001$, respectively, to a working point with $\eta_b=0.85$,
$p_c=0.1$, $p_j=0.01$.  In the case of the FCC-he configuration file,
we leave the default $b$-tagging working point unchanged which, for
$|y|<2.5$, $5<p_T<400$ GeV, is given by $\eta_b=0.85$, $p_c=0.04$,
$p_j=0.001$.

\begin{figure}[ht]
  \centering{}
   \includegraphics[scale=0.975]{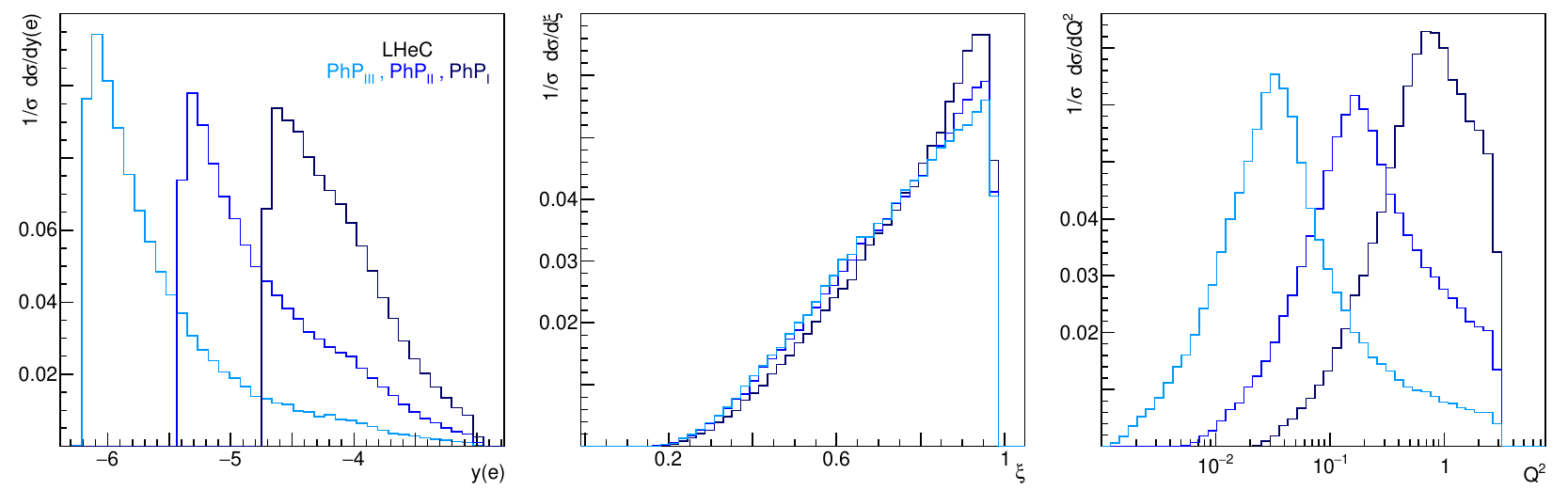}
   \caption{Normalized differential cross sections, with cuts
     $C_{0,\ldots,3}$, with respect to lab-frame scattered-electron
     rapidity (left panel), photon energy fraction (center) and photon
     squared virtuality (right). The three photoproduction regions
     (\ref{eq:php}) are illustrated, as indicated by the color code.}
  \label{fig:escat}
\end{figure}

We assume that the scattered electron in the photoproduction process
is detected either in the main detector, or in an appropriate forward
one.  We consider three rapidity intervals in which the scattered
electron can be detected, thus determining three photoproduction
regions.  First, we assume that the main detector covers the range
1$^\circ$--179$^\circ$ (see section 11 of \cite{LHeC12}),
corresponding to $y=-4.741$ in the backward hemisphere.  In fact, from
tables 12.4 and 12.6 of \cite{LHeC12} we see that both the tracker and
electromagnetic calorimeter are expected to extend up to $\eta=-4.8$,
so in this photoproduction region the scattered electron's
energy-momentum are measurable.  Second, we assume that the scattered
electron may be identified by a detector similar to the ones used for
Compton scattering and luminosity measurements, with angular
acceptance in the range 179$^\circ$--179.5$^\circ$, corresponding to
$-5.435< y < -4.741$.
Third, we assume that a small-angle electron tagger (see
\cite{LHeC12}, section 13.1.4) covers the range $\pi-4$ mrad
$<\theta(e^-)<\pi-8.727$ mrad, or $-6.215<y(e^-)<-5.435$. In summary,
we define the three photoproduction regions,
\begin{equation}
  \label{eq:php}
  \begin{aligned}
  PhP_{I}:   &\quad  -4.741< y(e^-) < -3.0,\\
  PhP_{II}:  &\quad  -5.435< y(e^-) < -3.0,\\
  PhP_{III}: &\quad  -6.215< y(e^-) < -3.0.
\end{aligned}
\end{equation}
For simplicity, we assume there are no gaps in each range.  Notice
that each region in (\ref{eq:php}) is defined to be contained by the
following one. In fact, the upper limit of the rapidity interval is
not very important in regions $PhP_{II}$ and $PhP_{III}$, as the
dominant contribution to the SM photoproduction cross section
originates in the lower part of the rapidity range, as illustrated in
figure \ref{fig:escat} (see left panel).

\begin{figure}[ht]
  \centering{}
   \includegraphics[scale=0.6]{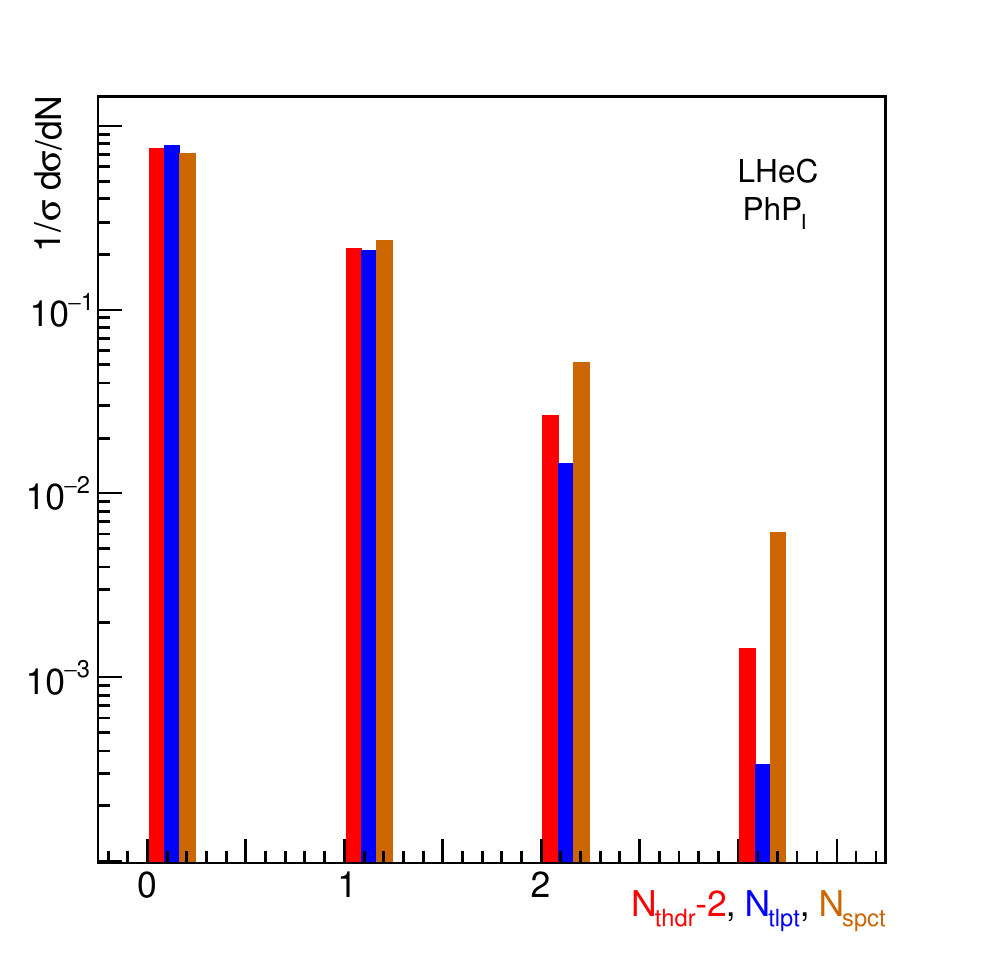}
   \caption{Normalized differential cross sections, with cuts
     $C_{0,\ldots,3}$, with respect to the number of light jets in
     hadronic top decay (red bars), in leptonic top decay (blue bars),
     and not associated with top decay (ocher bars).}
  \label{fig:njet}
\end{figure}

In order to determine the event kinematics we must first reconstruct
the neutrino momentum.  For processes such as $t\bar{t}$ and $tbW$
production in semileptonic mode, the transverse neutrino momentum is
identified with the missing transverse momentum, and its longitudinal
component is approximately reconstructed from $W$-decay kinematics.
In appendix \ref{sec:neutr.reco} we discuss in some detail our
approach to neutrino-momentum reconstruction in asymmetric colliders
such as the LHeC and FCC-he.  We identify the two $b$-tagged jets with
the highest $p_T$ as the ones originating in top decays, and denote
them $J_{b\,0,1}$. The remaining $b$-tagged jets, if any, are treated
as light jets.  The set of light jets is ordered by decreasing $p_T$,
and denoted $J_{k}$, $k=0,\ldots,N_l-1$. We identify the highest $p_T$
lepton, electron or muon, as the one arising from leptonic top decay.
With this information, we divide the light jets into three subsets:
jets arising from the hadronic top decay, $J_{\mathrm{hdr}\,k}$,
$k=0,\ldots,N_{t \mathrm{hdr}}-1$, jets from the leptonic top decay,
$J_{\mathrm{lpt}\,k}$, $k=0,\ldots,N_{t \mathrm{lpt}}-1$, and jets not
arising from top decay, or ``spectator'' jets, $J_{\mathrm{spct}\,k}$,
$k=0,\ldots,N_\mathrm{spct}-1$. Thus, the total number of jets in the
event is
$N_J=N_b+N_{t \textrm{hdr}}+N_{t \textrm{lpt}}+N_\textrm{spct}$.  Each
set of jets is ordered by decreasing $p_T$.  We introduce the $\chi^2$
function
  \begin{equation}
    \label{eq:chisq}
    \chi^2 = \frac{1}{\Gamma_t^2} \left(
      \left( m_{b_0\ell\nu j_\mathrm{lpt}} - m_t \right)^2 +
      \left( m_{b_1j_\mathrm{hdr}} - m_t \right)^2      
      \right),
\end{equation}
where $m_{b_0\ell\nu j_\mathrm{lpt}}$ is the invariant mass of one of
the two $b$-jets, together with the charged lepton and neutrino, and
with the jets $J_{\mathrm{lpt}}$. Similarly, $m_{b_1j_\mathrm{hdr}}$
is the mass of the other $b$-jet together with the jets
$J_\mathrm{hdr}$.  We set $m_t=172.5$ GeV and $\Gamma_t=1.42$ GeV
\cite{PDG}.  Out of all possible jet configurations, the one that
minimizes $\chi^2$, as given in (\ref{eq:chisq}), is selected, leading
to reconstructed top and antitop quarks.  The leptonic $b$-jet is
denoted $J_{b\,0}$ and the hadronic one $J_{b\,1}$.  The distribution
of the number of light jets originating from hadronic and leptonic top
decays, and not arising from top decays is displayed in figure
\ref{fig:njet} at the LHeC energy in photoproduction region $PhP_I$.
In the other regions it is similar, and at the FCC-he energy a modest
increase in the number of light jets is observed, especially in
$N_{t \textrm{hdr}}$. The event is retained if this reconstructed
kinematics satisfies the phase-space cuts we discuss next.

In a top-pair production event in semileptonic channel for which
(\ref{eq:chisq}) can be defined there must be at least two $b$ jets
and at least two light ones.  We then keep only the events
satisfying the preselection cut
\begin{equation}
  \label{eq:pre.sel}
  2\leq N_b~, \qquad 2\leq N_l \leq 5~.
\end{equation}
Here the requirement of at most five light jets is applied for
computational simplicity since, out of the events with $N_l\geq 2$,
approximately 98.5\% have $N_l\leq 5$.  Events fulfilling
(\ref{eq:pre.sel}) can be reconstructed using (\ref{eq:chisq}).

We introduce a phase-space cut defining the photoproduction kinematic
region, 
\begin{equation}
  \label{eq:C0}
  C_0:\quad\left\{
  \begin{gathered}
    y(e^-) \in PhP_{k},\quad k=I,II,III,\\
    Q^2 < 3\,\mathrm{GeV}^2,\; 0<\xi<0.98,\\
    \Delta R(j,j) > 0.4,    
  \end{gathered}
  \right.
\end{equation}
where $e^-$ refers to the scattered electron, and the first line
refers to the appropriate photoproduction region (\ref{eq:php}). In
(\ref{eq:C0}), $Q^2=-q_\gamma^2$ refers to the quasireal photon's
squared virtuality, and $\xi=q_\gamma^0/E_{e\mathrm{beam}}$ to the
photon's fraction of the electron-beam energy in the lab frame.  We
assume that the scattered electron must have an energy of at least 1.2
GeV to be detected at the very forward detector, which leads to the
upper cut in $\xi$ in (\ref{eq:C0}).  The distributions of the
scattered-electron lab-frame rapidity, $y(e^-)$, and the photon
energy fraction $\xi$ and $Q^2$ are shown in figure \ref{fig:escat} in
the three photoproduction regions (\ref{eq:php}) at the LHeC energy.
$\Delta R(j,j)$ in (\ref{eq:C0}) refers to the distance between any
pair of jets, $\Delta R=\sqrt{(\Delta\eta)^2+(\Delta\varphi)^2}$.
\begin{table}[t]
  \centering
  \begin{tabular}{cccc|ccc}
    & \multicolumn{3}{c|}{LHeC} & \multicolumn{3}{c}{FCC-he} \\
    & $PhP_{I}$ & $PhP_{II}$ & $PhP_{III}$& $PhP_{I}$ & $PhP_{II}$ & $PhP_{III}$ \\\hline
    $p^\perp_{j\,\textrm{min}}$ [GeV]& 10 & 15 & 20 & 15 & 20 & 25\\
    $y_{b,\mathrm{min}}$&-0.5&-0.5&-0.5&-0.5&-0.5&-0.5\\
    $y_{b,\mathrm{max}}$&3.0&3.0&3.0&3.0&2.7&2.7\\
    $p^\perp_{b,\mathrm{min}}$ [GeV]&30&35&45&35&40&50\\
    $p^\perp_{b,\mathrm{max}}$ [GeV]&$\infty$&$\infty$&$\infty$&$\infty$&200&200\\
  \end{tabular}
  \caption{Parameters involved in the cuts $C_0$,\ldots,$C_4$, equations (\ref{eq:C0}),\ldots,(\ref{eq:C4}).}
  \label{tab:tab1}  
\end{table}

\begin{figure}
  \centering{}
\includegraphics[scale=0.95]{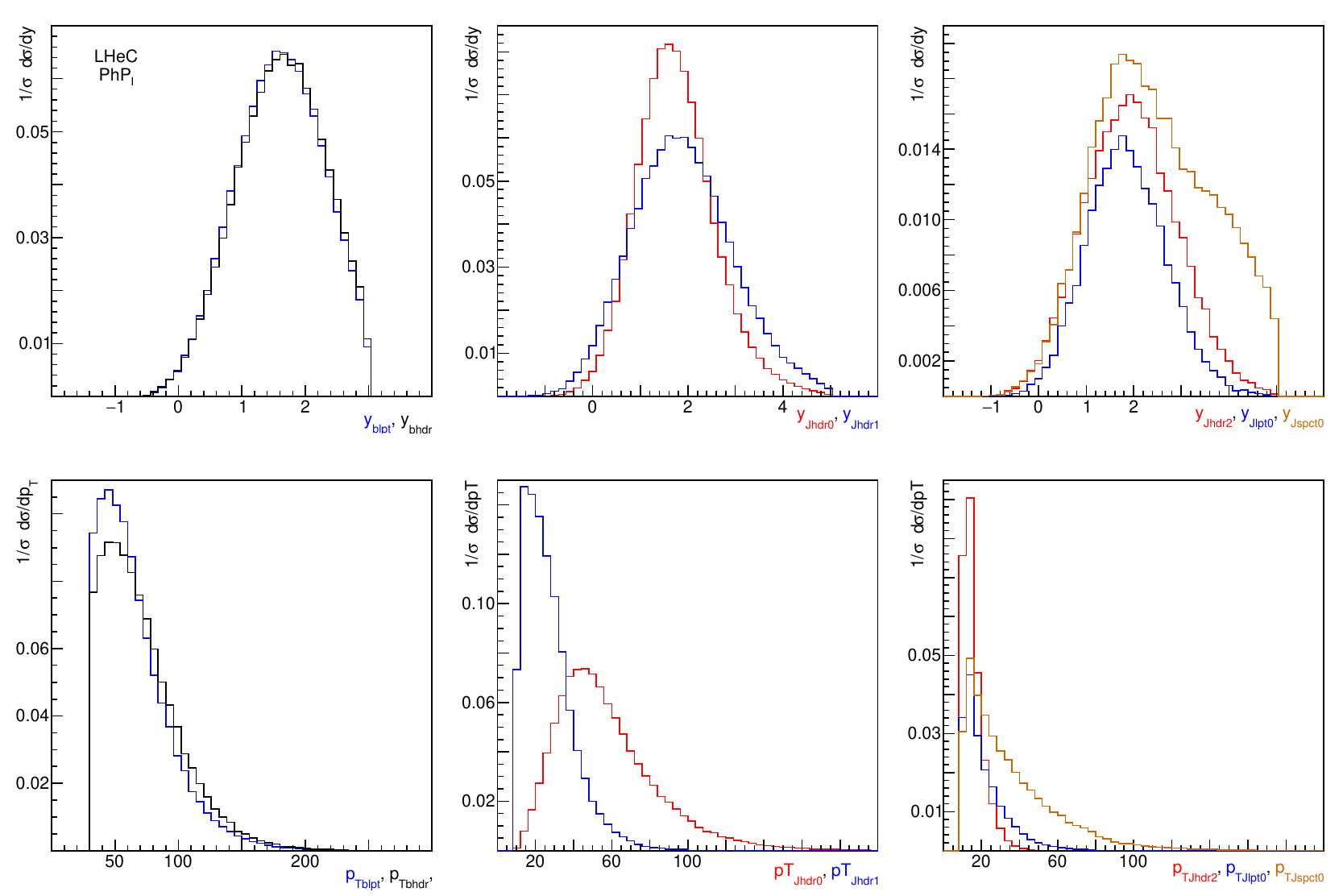}  
\caption{Normalized differential cross sections, with cuts
  $C_{0,\ldots,3}$, with respect to lab-frame rapidity (upper row) and
  transverse momentum (lower row). Left column: $b$ jet from leptonic,
  $J_{b0}$, and hadronic, $J_{b1}$, top decay.  Central column: two
  leading-$p_T$ light jets from hadronic top decay. Right column:
  third leading-$p_T$ light jet from hadronic, and leading $p_T$ jet
  from leptonic top decay, and leading jet not originating in top
  decay.}
  \label{fig:c8}
\end{figure}

We identify the charged lepton $\ell$ from top decay with the
leading-$p_T$ lepton.  Both the LHeC and FCC-he detectors have a
rapidity acceptance $|y|<4$ for muons.  We use a cut for the charged
lepton that essentially reflects that acceptance window,
\begin{equation}
  \label{eq:C1}
  C_1:\quad -2.8<y(\ell)<4, \quad |\vec{p}^\perp(\ell)|> 5.0\,\mathrm{GeV}, 
  \quad |\vec{p}^\perp_\mathrm{miss}|> 5.0\,\mathrm{GeV}.
\end{equation}
We set a rapidity cut on all light jets that roughly follows the
upper limit of the rapidity acceptance range for both the LHeC
($y<4.9$) and FCC-he ($y<5.2$) tracking systems.  We identify the two
leading $p_T$ light jets as originating from top decay, and we require
them to satisfy a minimum $p_T$ cut.
\begin{equation}
  \label{eq:C2}
  C_2:\quad -2.0 < y(j)< 5.0,
  \quad |\vec{p}^\perp(J_{\mathrm{hdr}\,
    0})|,|\vec{p}^\perp(J_{\mathrm{hdr}\,1})|> 
  p^\perp_{j\,\textrm{min}}. 
\end{equation}
Here $j$ refers to all jets except the two $b$-tagged ones with the
highest transverse momenta.  $J_{t_\mathrm{hdr} k}$ refers to the
$p_T$-ordered collection of jets involved in the hadronic top
decay ($k=0$, 1,\ldots), as defined above.  Eq.\ (\ref{eq:C2}) requires the two hardest
such ones to have $p_T>p^\perp_{j\,\mathrm{min}}$ as given in table
\ref{tab:tab1}.  The remaining light jets $J_{\mathrm{hdr}\, k}$ with
$k>1$, as well as the light jets $J_{\mathrm{lpt}\,k}$ arising from
the leptonic decay and those not originating from a top decay, are
required to satisfy $|\vec{p}^\perp|> 10$  GeV. The differential cross
sections with respect to light jet lab-frame rapidity and transverse
momentum is shown in figure \ref{fig:c8}. The center-column panels in
that figure refer to the two leading-$p_T$ jets originating in
hadronic top decay.  The right-column panels show the distributions
for the leading supernumerary jets, $J_{\mathrm{hdr}\,2}$,
$J_{\mathrm{lpt}\,0}$, $J_{\mathrm{spct}\,0}$.  Notice that in this
last case the distributions have a normalization deficit, since not
all events contain supernumerary jets. 

As discussed in connection with the reconstruction of the top quarks,
we identify the $b$ jets produced directly in the top decays with the
two leading-$p_T$ $b$-tagged jets.  The cuts in these jets' kinematic
variables are crucial to suppress the $tbW$ irreducible background.
We adopt the following cuts,
\begin{equation}
  \label{eq:C3}
  C_3:\quad\left\{
    \begin{gathered}
      y_{b,\mathrm{min}}<y(J_{bk})<y_{b,\mathrm{max}}\\
      p^\perp_{b,\mathrm{min}} < |\vec{p}^\perp(J_{bk})| <
      p^\perp_{b,\mathrm{max}} 
    \end{gathered}\right.,
  \quad k=0,1,
\end{equation}
where the relevant parameters are given in table \ref{tab:tab1}.  The
parameter ranges in (\ref{eq:C3}) are chosen so as to suppress the
$tbW$ background as much as possible, but wide enough for the cuts to
remain generic.

\begin{figure}
  \centering{}
\includegraphics[scale=0.95]{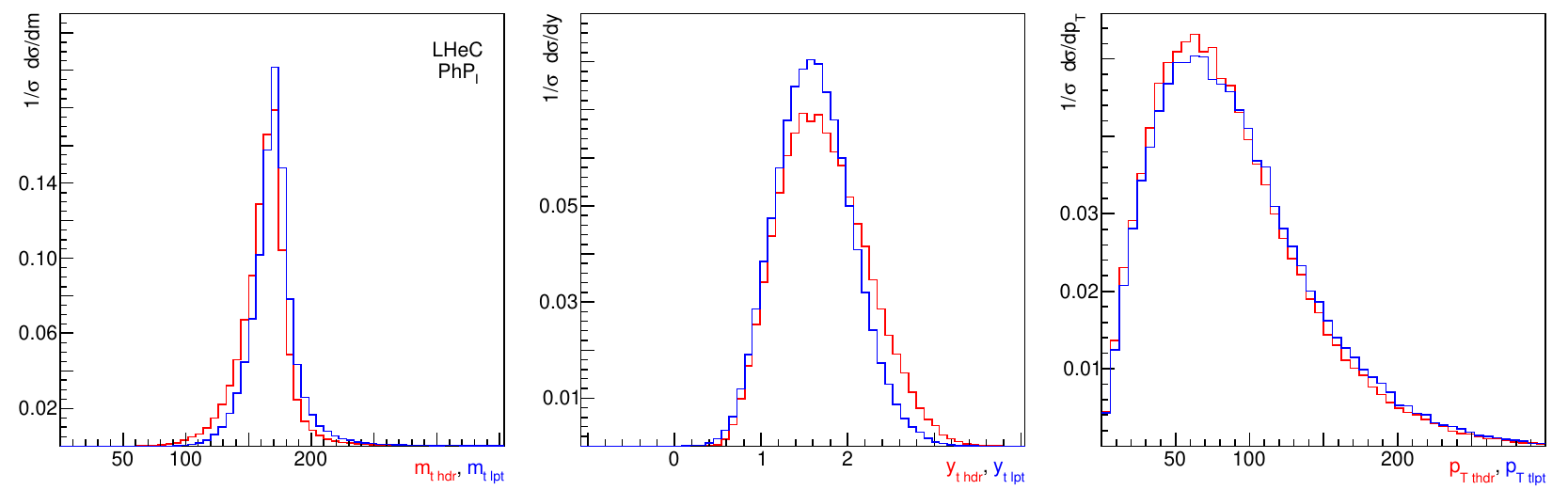}  
  \caption{Normalized differential cross sections, with cuts
     $C_{0,\ldots,3}$, with respect to hadronically and leptonically
     decaying top quark mass (left panel), lab-frame rapidity (center
     panel) and transverse momentum (right panel). 
}
  \label{fig:c9}
\end{figure}

We introduce also a cut on the top masses,
\begin{equation}
  \label{eq:C4}
  C_4:\quad
  \left|m_{t\,\mathrm{hdr}}-m_t\right|^2 +
  \left|m_{t\,\mathrm{lpt}}-m_t\right|^2 < (\Delta m_t)^2,
\end{equation}
where $m_{t\,\mathrm{hdr}}$, $m_{t\,\mathrm{lpt}}$ refer to the
reconstructed masses of the hadronically and leptonically decaying top
quarks, and where we take $m_t=172.5$ GeV, as already mentioned, and
$\Delta m_t=30$ GeV.  The mass distributions for the hadronically and
leptonically decaying top quarks are shown in figure \ref{fig:c9} (see
the left panel).  Also shown in the figure are the lab-frame rapidity
and transverse momentum distributions for the hadronic and leptonic
top quarks.

Once the cuts $C_{0\text{--}4}$ have been applied, the cross sections
for the $t\bar{t}$ photoproduction signal process (\ref{eq:ttx.proc}),
figure \ref{fig:feyn.1}, and the $tbW$ irreducible background
(\ref{eq:tbw.proc}), figure \ref{fig:feyn.2}, are found to be as
follows,
\begin{equation}
  \label{eq:sm.xsctn.numeric}
  \begin{array}{c|ccc|ccc|}
             &\multicolumn{3}{c|}{\text{LHeC}}&\multicolumn{3}{c|}{\text{FCC-he}}\\
{[}\text{fb}]  &PhP_{I}&PhP_{II}&PhP_{III}&PhP_{I}&PhP_{II}&PhP_{III}\\\hline
t\bar{t} &0.40 &0.73 &1.32&4.28&6.19&10.51\\
tbW      &0.041&0.083&0.16&0.44&0.71&1.42
  \end{array}~,
\end{equation}
expressed in femtobarns.  We notice here that the $tbW$ background has
cross section at the parton level that is roughly 20\% of the signal
cross section at the LHeC, and roughly 35\% at the FCC-he, the precise
number depending on the photoproduction region. We designed the cuts
$C_{0,\ldots,4}$ to reduce this background to levels below 15\%.  As
seen in (\ref{eq:sm.xsctn.numeric}), the $tbW$ background is 10\% of
the signal in region $PhP_{I}$ and 11.3\% in $PhP_{II}$ at both the
LHeC and FCC-he. In region $PhP_{III}$ we have 12.2\% at the LHeC and
13.5\% at the FCC-he.  The $tbW$ background proves to be the most
difficult one to control.  The irreducible background $tbqq$
(\ref{eq:tbqq.proc}), figure \ref{fig:feyn.3}, leads to cross sections
at the parton level that are at most 0.5\% of those of the signal
process $t\bar{t}$.  At the detector level, when the cuts
$C_{0,\ldots,4}$ are applied, these cross sections fall below the
0.1\% level.  We therefore neglect this background in what follows.

\section{Further SM backgrounds}
\label{sec:sm.bck}

In the previous section the signal process (\ref{eq:ttx.proc}) and two
of its irreducible backgrounds, (\ref{eq:tbw.proc}) and
(\ref{eq:tbqq.proc}), both involving resonant top production, were
discussed in detail.  In this section we discuss several additional
irreducible and reducible SM backgrounds. The $tbW$ production process
(\ref{eq:tbw.proc}) turns out to be the most important background,
followed by single-top $tbq$ production (section \ref{sec:tbq}), $Whq$
production (section \ref{sec:notop.q}), and $tbW$ again, in its
reducible version (\ref{sec:tbw.red}). A summary of their cross
sections relative to the signal process is given below in table
\ref{tab:summary}.

\subsection{Single-top photoproduction ($tbq$)}
\label{sec:tbq}

The single-top, or $tbq$, photoproduction irreducible background is of
the form,
\begin{equation}
  \label{eq:tbq.1}
  q e^- \rightarrow e^- t \bar{b} q' + e^- \bar{t} b q'
  \rightarrow e^-\, b\ell^+\nu_\ell \,\bar{b} q' + e^-\, \bar{b}
  \ell^- \bar{\nu}_\ell \, b q'~.
\end{equation}
With all possible lepton- and quark- flavor combinations taken into
account, this process is given by 320 Feynman diagrams. For this
process to pass the preselection cut (\ref{eq:pre.sel}), an
additional parton must be radiated.  We may obtain that additional jet
in the final state by QCD showering, or by considering the
photoproduction of the final states (\ref{eq:tbq.1}) with an
additional gluon. For the purpose of obtaining the cross section,
either process, or an appropriately matched combination of the two,
leads to the same result.

The process of $tbqg$ photoproduction is determined by 1800 Feynman
diagrams distributed among 32 flavor channels. The cross section for
(\ref{eq:tbq.1}), however, is dominated by subprocesses with an
initial valence quark undergoing a Cabibbo-allowed transition.  In
particular, the cross section for final states with an antilepton
$\ell^+$ is twice as large as that for final states with $\ell^-$,
since the former involve $u\rightarrow d$, while the latter involve
$d\rightarrow u$.  Judicious choice of the 10 flavor channels with the
largest cross sections leads to a reduction of the number of diagrams
to 585 without loss of numerical precision.  From the diagrams for
$tbqq$ photoproduction in figure \ref{fig:feyn.3} we can obtain
diagrams for $tbqg$ by crossing the initial gluon to the final state,
and one final-state light quark to the initial state.  Further
diagrams can be obtained from those by reattaching the final gluon to
other colored lines in the diagram.

Full simulation, including the cuts $C_{0\text{--}4}$, leads to the
cross sections for this process,
\begin{equation}
  \label{eq:tbq.2}
  \begin{array}{c|ccc|ccc|}
             &\multicolumn{3}{c|}{\text{LHeC}}&\multicolumn{3}{c|}{\text{FCC-he}}\\
{[}\text{fb}]  &PhP_{I}&PhP_{II}&PhP_{III}&PhP_{I}&PhP_{II}&PhP_{III}\\\hline
tbq  &0.014&0.032&0.061&0.039&0.072&0.13\\
tbqg &0.017&0.029&0.046&0.041&0.060&0.091                      
  \end{array}~,
\end{equation}
expressed in femtobarns.  We see that this background constitutes 4\%
of the signal cross section at the LHeC, and 1\% at the FCC-he.

\subsection{No-top irreducible background: gluon-initiated processes}
\label{sec:notop.g}

In this section we consider irreducible backgrounds to the signal
process (\ref{eq:ttx.proc}), described by Feynman diagrams not
containing top lines.  We begin by discussing the processes with
$ge^-$ initial states; the ones with initial $qe^-$, with $q$ a light
quark, are the topic of the next subsection.

\begin{figure}[ht]
  \centering{}
   \includegraphics[scale=1.0]{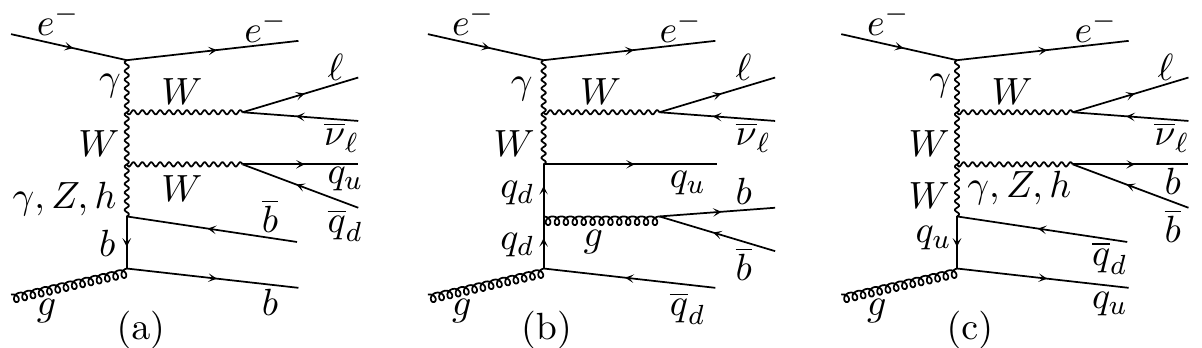}
  \caption{Representative unitary-gauge Feynman diagrams for the
    photoproduction of (a) $WWb\bar{b}$ in semileptonic mode,
    (b) $Wg^*qq$, (c) $Whqq$, $WZqq$, $W\gamma^*qq$. These diagrams
    belong to the set QCD$^1_{bbg}$, QCD$^3$, QCD$^1_{qqg}$
    respectively, see text.}
  \label{fig:notop.g}
\end{figure}

We consider processes with the same initial and final state as the
signal one (\ref{eq:ttx.proc}),
\begin{equation}
  \label{eq:notop.g.1}
ge^-\rightarrow e^-\, b\ell^+\nu_\ell\,
    \bar{b}\bar{q}_uq_d + e^-\, bq_u\bar{q}_d\,
    \bar{b}\ell^-\bar{\nu}_\ell~,  
\end{equation}
but described by Feynman diagrams without internal top lines.  Taking
into account the four quark-flavor channels and the duplication of the
number of diagrams in the $e^-$ lepton channel, there are 7920
diagrams for the process (\ref{eq:notop.g.1}), without $t$ lines and
with one QCD vertex, in the photoproduction regime. We denote that set
of diagrams QCD$^1$.  There is also a set of 1880 diagrams without $t$
lines and with three QCD vertices, which we denote QCD$^3$.  Some
diagrams for the process (\ref{eq:notop.g.1}) are illustrated in
figure \ref{fig:notop.g}.

We consider first the diagrams with only one QCD vertex, set QCD$^1$.
We divide that set into two subsets,
$\mathrm{QCD}^1=\mathrm{QCD}^1_{bbg}\cup\mathrm{QCD}^1_{qqg}$, with
QCD$^1_{bbg}$ the set of diagrams in which the QCD vertex is $bbg$,
and QCD$^1_{qqg}$ that of diagrams with a $qqg$ vertex, with $q$ any
light quark. The set QCD$^1_{bbg}$ contains 3240 diagrams, and
QCD$^1_{qqg}$ contains 4680 diagrams.  The set QCD$^1_{bbg}$ has a
cross section that is between 70\% and 90\% of the cross section for
the entire set QCD$^1$, depending on the energy and photoproduction
region, with the set QCD$^1_{qqg}$ providing the rest of the cross
section.

The cross section obtained from the set QCD$^1_{bbg}$ is dominated by
the contribution from the diagrams for the process
(\ref{eq:notop.g.1}) with $WWbb$ intermediate state,
$ge^- \rightarrow e^- b\overline{b}\, W^+ W^-$, as illustrated in
figure \ref{fig:notop.g} (a). The photoproduction of $WWbb$ is given
by 720 diagrams and yields more than 90\% of the QCD$^1_{bbg}$ cross
section. The cross sections for QCD$^1_{bbg}$ are small relative to
the signal process already at the parton level, with the largest
percentage fraction obtained at the FCC-he energy and in the
$PhP_{III}$ region, and found to be 3\%.  Even in that case however,
at the detector-simulation level and with the cuts from section
\ref{sec:php.mc}, the cross section for QCD$^1_{bbg}$ is found to
amount to 0.08\% of the signal process, so we consider it negligible.

The set of diagrams QCD$^1_{qqg}$ gives a smaller fraction of the
cross section for the entire set QCD$^1$, representing from 20\% in
the region $PhP_I$ to 30\% in $PhP_{III}$ at the LHeC energy, and
about 12\% at the FCC-he. The main contribution to the cross section
from this set originates in the doubly resonant production of $hWqq$,
followed by the leptonic decay of $W$ and $h\rightarrow b\bar{b}$.
This process is given by 320 Feynman diagrams, with a representative
example shown in figure \ref{fig:notop.g} (c).  A smaller contribution
to the cross section originates in the photoproduction of $ZWqq$, also
shown in the figure.  This process is given by 1000 Feynman diagrams.
Its cross section is at most one-third of that for $hWqq$ (in $PhP_I$
at the LHeC) to less than 1\% of $hWqq$ at the FCC-he.  There is a set
of less resonant processes similar to $hWqq$, such as $h\ell\nu qq$
and $Z\ell\nu qq$ photoproduction, that yield much smaller cross
sections.  The remaining diagrams in the set QCD$^1_{qqg}$ not
considered so far involve the vertex $\gamma^*\rightarrow b\bar{b}$.
These are 2040 diagrams that yield a cross section much smaller than
$hWqq$ photoproduction, and can therefore be neglected.  Just as with
the processes described by QCD$^1_{bbg}$, the contribution of
QCD$^1_{qqg}$ to the cross section is negligibly small compared to the
signal cross section.

The set of diagrams QCD$^3$, illustrated in figure \ref{fig:notop.g}
(b), is at most singly resonant and leads to small corrections to the
cross sections obtained from the set QCD$^1$, so we will not consider
it further here.  We mention, however, that this set of diagrams is
closely related to the reducible background discussed below in section
\ref{sec:red.bck.g}.  As an illustration of this fact we notice that
diagram \ref{fig:notop.g} (b) is the same as \ref{fig:red.bck.g} (b)
with $q=b$.  Thus, the cross section for the QCD$^3$ processes is
actually taken into account when showering the backgrounds in section
\ref{sec:red.bck.g}.

\subsection{No-top irreducible background: quark-initiated processes}
\label{sec:notop.q}

\begin{figure}[ht]
  \centering{}
   \includegraphics[scale=0.975]{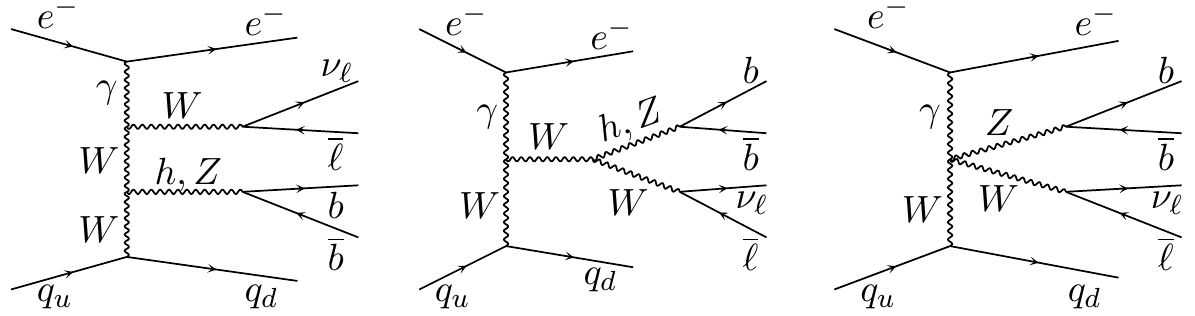}
  \caption{Sample unitary-gauge Feynman diagrams for the
    photoproduction of $Whq$, $WZq$ in semileptonic mode.}
  \label{fig:notop.q}
\end{figure}

In this section we turn our attention to processes with one less light
parton in the final state as the signal one (\ref{eq:ttx.proc}),
\begin{equation}
  \label{eq:notop.q.1}
qe^-\rightarrow e^- q b\bar{b}\ell\nu_\ell,  
\end{equation}
and described by Feynman diagrams without internal top lines.  Taking
into account the four quark-flavor channels and the duplication of the
number of diagrams in the $e^-$ lepton channel, there are
3320 diagrams for process
(\ref{eq:notop.q.1}), without $t$ lines, at lowest QCD order and in
the photoproduction regime.  A few of those diagrams are illustrated
in figure \ref{fig:notop.q}.
The cross section for the processes (\ref{eq:notop.q.1}) is strongly
dominated by $Whq$, and to a lesser extent $WZq$, associated
photoproduction as illustrated in the figure.  In fact, these doubly
resonant processes account for about 97\% of the cross section for
(\ref{eq:notop.q.1}), with $Whq$ alone yielding 90\% of the total.

If we restrict ourselves to those doubly resonant processes, the
number of diagrams is reduced to 36 per flavor channel.  However, in
order to compute the cross section, it is enough to consider processes
with a valence quark in the initial state and a Cabibbo-allowed flavor
transition, and only one lepton flavor,
\begin{equation}
  \label{eq:notop.q.2}
  ue^-\rightarrow e^- d b\bar{b}\bar{\mu}\nu_\mu,
  \quad
  de^-\rightarrow e^- u b\bar{b}\mu\bar{\nu}_\mu,  
\end{equation}
and multiply the cross section $\times2.11$ at the LHeC and
$\times2.28$ at the FCC-he.

The numerical results for the cross section for this irreducible
background are as follows,
\begin{equation}
  \label{eq:notop.1}
  \begin{array}{c|ccc|ccc|}
             &\multicolumn{3}{c|}{\text{LHeC}}&\multicolumn{3}{c|}{\text{FCC-he}}\\
{[}\text{fb}]  &PhP_{I}&PhP_{II}&PhP_{III}&PhP_{I}&PhP_{II}&PhP_{III}\\\hline
Whq+WZq  &0.007&0.014&0.018&0.011&0.016&0.013
  \end{array}~.
\end{equation}
These cross sections are rather small, though not negligible, at the
LHeC, where they constitute 1.8\% of the signal cross section after
cuts, but much less significant at the FCC-he where they are 0.2\% of
the signal.

\subsection{Photoproduction of $tbW$ as reducible background}
\label{sec:tbw.red}

In the LHeC simulation in Delphes 3.4.2 with default parameters, the
$b$-tagging algorithm only operates on jets with lab.-frame absolute
rapidity $|y(j)|<3$, and $p_T>0.5$ GeV.  In the case of the FCC-he
detector simulation, $b$ tagging is restricted to $|y(j)|<4$, and
$p_T>4.0$ GeV. It is then possible to have $tbW$ events in which one
$b$-jet with $|y(j_b)|<y_\mathrm{max}=3$, 4 is tagged and passes the cuts (\ref{eq:C3}),
whereas the other $b$-jet has $y(j_b)>3$ and is therefore not tagged,
but passes the cuts (\ref{eq:C2}) for light jets.  Under those
conditions, if a light jet is mistagged and satisfies the conditions
(\ref{eq:C3}), the event could pass the cuts.  We conclude, then, that
the process $tbW$ has a reducible component that we must consider.
We estimate the cross section for this process by requiring at the
parton level that one $b$ quark be inside the flavor-tagging region and
the other one outside.  The cross sections we obtain are
\begin{equation}
  \label{eq:tbw.red}
  \begin{array}{c|ccc|ccc|}
             &\multicolumn{3}{c|}{\text{LHeC}}&\multicolumn{3}{c|}{\text{FCC-he}}\\
{[}\text{fb}]  &PhP_{I}&PhP_{II}&PhP_{III}&PhP_{I}&PhP_{II}&PhP_{III}\\\hline
tbW_\mathrm{red}  &0.004&0.006&0.013&0.008&0.007&0.005
  \end{array}~.
\end{equation}
We conclude that this reducible component of the $tbW$ background
constitutes about 1\% of the signal cross section at the LHeC in all
three photoproduction regions, and not more than 0.2\% at the FCC-he.

\subsection{Reducible backgrounds: quark-initiated processes}
\label{sec:red.bck.q}

\begin{figure}
  \centering{}
  \includegraphics[scale=0.975]{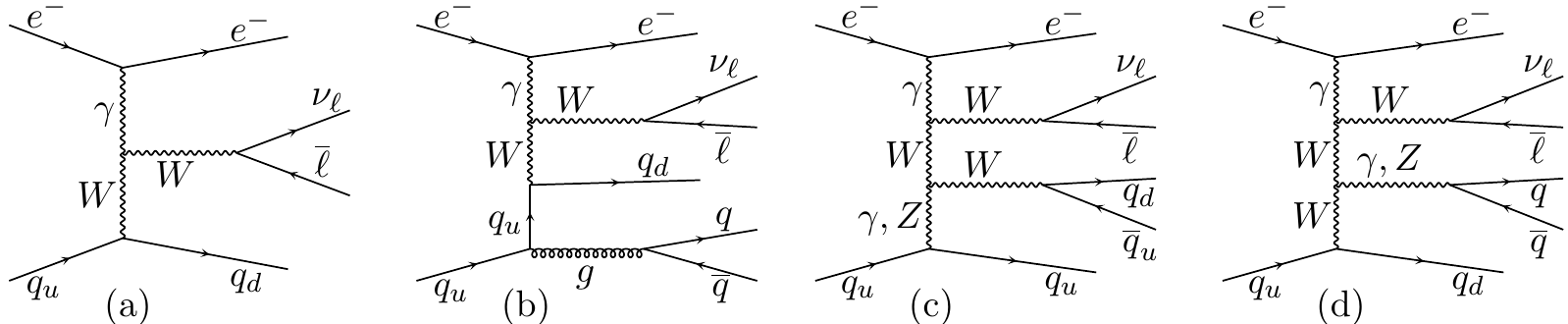}
  \caption{Sample unitary-gauge Feynman diagrams for quark-initiated
    reducible background processes including photoproduction of (a) $Wq$,
  (b) $Wqqq$, (c) $WWq$, (d) $WZq$/$W\gamma^*q$.}
\label{fig:red.bck.q}  
\end{figure}
The quark-initiated photoproduction processes with one charged lepton,
missing transverse energy and jets in the final state are described by
Feynman diagrams with $N_p=N_g+N_q$ partons in the final state, where
$N_g\geq0$ is the number of gluons and $N_q\geq1$ odd is the number of
quarks, and exactly $N_{QCD}$ strong and ($N_p+3-N_{QCD}$) electroweak
vertices.  We will restrict ourselves to $N_q=1,$ 3, since larger
values lead to processes with very small cross sections.

For $N_q=1$ we obtain leptonic single-$W$ production associated with at
least one jet, as illustrated by the Feynman diagram in figures
\ref{fig:red.bck.q} (a), (b).  For the purpose of estimating the cross
section we take into account at the parton level only diagrams with
$N_{QCD}=0$, such as \ref{fig:red.bck.q} (a), and deal with QCD
radiation only through Pythia showering.\footnote{Therefore, we cannot
  exclude contributions such as that in figure \ref{fig:red.bck.q} (b)
  with $q=b$, which constitute an irreducible background.  We could
  argue about the consistency of including those here, but that
  discussion is made irrelevant by the fact that all those
  contributions turn out to be negligible, as shown below.}  Thus,
when all possible lepton- and quark- flavor combinations are allowed
for, we have 160 diagrams analogous to that of figure
\ref{fig:red.bck.q} (a) with exactly one final-state quark and four
electroweak vertices.  The cross sections are quite small, as this
background represents 0.2\% of the signal process at the LHeC, region
$PhP_I$, and less than 0.1\% in the other regions.  At the FCC-he this
background essentially vanishes, due to the higher purity of the
$b$-tagging algorithm.

For $N_q=3$, at lowest order in QCD, we have diagrams with three
final-state quarks and six electroweak vertices as shown in figures
\ref{fig:red.bck.q} (c) and (d).  There are 40500 such Feynman
diagrams when all 320 possible quark- and lepton- flavor combinations
are taken into account.  Out of the many physical processes
contributing to the cross section for this background, the dominant
ones are the doubly resonant $WW$ production, illustrated in figure
\ref{fig:red.bck.q} (c), and $WZ$ production, shown in
\ref{fig:red.bck.q} (d) together with the singly resonant production
of $W\gamma^*$.  Numerically, the cross sections for these processes
are typically a few percent of those for the $N_q=1$ processes,
therefore negligibly small. These results also strongly suggest that
we do not need to consider reducible backgrounds with $N_q\geq5$.

\subsection{Reducible backgrounds: gluon-initiated processes}
\label{sec:red.bck.g}

The gluon-initiated photoproduction processes with one charged lepton,
missing transverse energy and jets in the final state are described by
Feynman diagrams with $N_p=N_g+N_q$ partons in the final state, where
$N_g\geq0$ is the number of gluons and $N_q\geq2$ even is the number
of quarks, and exactly $N_{QCD}$ strong and ($N_p+3-N_{QCD}$)
electroweak vertices.  We will restrict ourselves to $N_q=2,$ 4, since
larger values lead to processes with very small cross sections.

\begin{figure}
  \centering{}
  \includegraphics[scale=1.0]{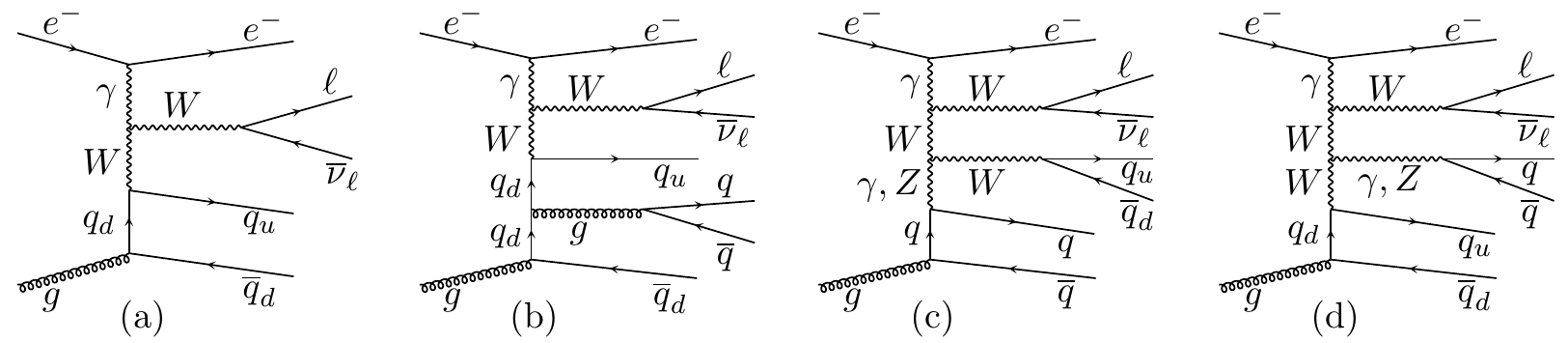}
  \caption{Sample unitary-gauge Feynman diagrams for gluon-initiated
    reducible background processes including photoproduction of (a) $Wqq$,
 (b) Wqqqq, (c) $WWqq$, (d) $WZqq$/$W\gamma^*qq$.}
\label{fig:red.bck.g}  
\end{figure}

For $N_q=2$ we obtain leptonic single-$W$ production associated with at
least two jets, as illustrated by the Feynman diagram in figure
\ref{fig:red.bck.g} (a), (b).  As in the case of quark-initiated
processes, we allow for further QCD radiation only through Pythia
showering. Thus, when all possible lepton- and quark-
flavor combinations are considered, we have 200 diagrams analogous to
that of figure \ref{fig:red.bck.g} (a) with exactly two final-state
quarks, one strong and four electroweak vertices.
The cross sections obtained for these processes with the cuts
described in section \ref{sec:php.mc} are negligibly small.  This
background represents less than 0.1\% of the signal process at the
LHeC, region $PhP_I$, and much less than that in the other regions.
At the FCC-he this background essentially vanishes, due to the higher
purity of the $b$-tagging algorithm.

For $N_q=4$, at lowest order in QCD, we have diagrams with four
final-state quarks, one strong and six electroweak vertices as shown
in figures \ref{fig:red.bck.g} (c) and (d).  There are 64800 such
Feynman diagrams when all 96 possible quark- and lepton- flavor
combinations are taken into account.  Out of the many physical
processes contributing to the cross section for this background, the
dominant ones are the doubly resonant $WW$ production, illustrated in
figure \ref{fig:red.bck.g} (c), and $WZ$ production, shown in
\ref{fig:red.bck.g} (d) together with the singly resonant production
of $W\gamma^*$.  Numerically, the parton-level cross sections for
these processes are typically less than 1\% of those for the
$N_q=2$ processes.  Thus, the contributions of these processes to the
background cross section are negligibly small. These results also
strongly suggest that we do not need to consider reducible backgrounds
with $N_q\geq6$.

\section{Results for effective couplings}
\label{sec:res.coupl}

For the computation of the cross section as a function of the
effective couplings $\cub$, $\cmpq$, we simulated the top-pair
photoproduction process in $pe^-$ collisions with MG5, Pythia
6, and Delphes 3 as described in section \ref{sec:php.mc}.  The
effective operators (\ref{eq:operators}) were implemented in MagGraph
5 by means of the program FeynRules version 2.0.33 \cite{all14}.

\begin{figure}
  \centering{}
  \includegraphics[scale=0.75]{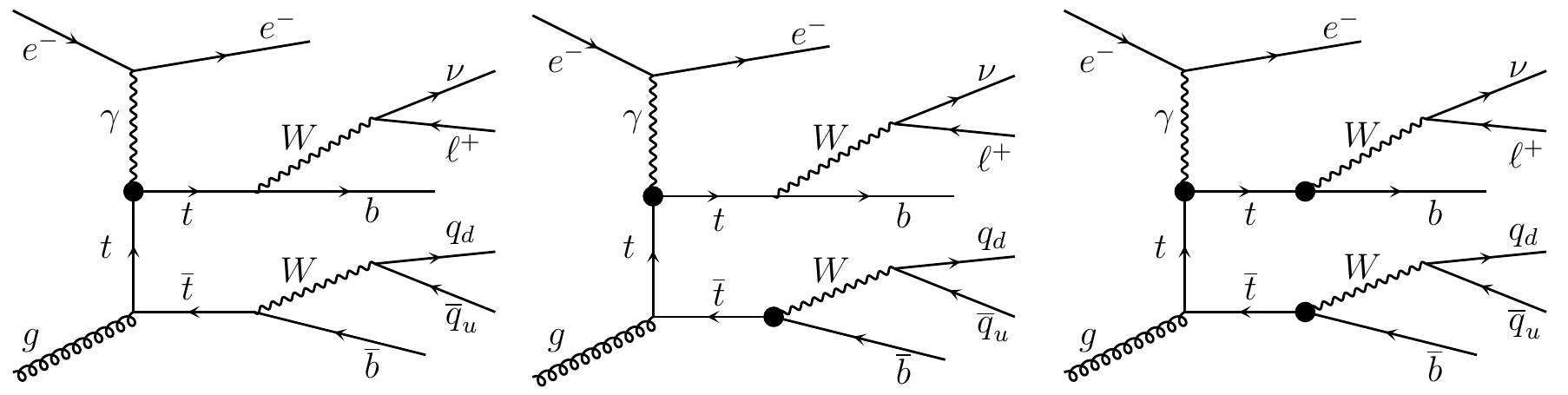}
  \caption{Representative unitary-gauge Feynman diagrams with one,
    two and three anomalous effective vertices.}
  \label{fig:feyn.zz}
\end{figure}

As discussed above in section \ref{sec:php.sm}, there are 180 diagrams for
semileptonic top-pair photoproduction and decay in the SM with Cabibbo
mixing, allowing for all possible quark- and (light) lepton- flavor
combinations. When the effective operators are included, additional
diagrams enter the computation.  If we include only the operator
$Q_{uB}^{33}$, there are 40 additional diagrams with one effective
vertex, for a total of 220 diagrams.  If we include only the operator
$Q_{\varphi Q}^{(-)33}$, there are 360 additional diagrams with one
effective vertex and 180 with two such vertices, for a total of 720
diagrams.  Finally, if we take into account both operators, then there
are 400 diagrams with one effective vertex, 260 with two and 40 with
three effective vertices, for a total of 880 diagrams.  Representative
diagrams with one, two and three effective vertices are shown in
figure \ref{fig:feyn.zz}.

Diagrams with one, two and three effective vertices entering the
amplitude for (\ref{eq:ttx.proc}), contribute to it at
$\mathcal{O}(\Lambda^{-n})$ with $n=2,$ 4 and 6, respectively. In
fact, once the top propagator dependence on effective couplings
through the top decay width is taken into account, the scattering
amplitude is given as a power series of $\Lambda^{-2}$.  We remark
that diagrams with two effective vertices must be kept in the
amplitude since, through their interference with SM diagrams, they
make contributions to the cross section of the same order,
$\mathcal{O}(\Lambda^{-4})$, as the square of diagrams with only one
effective vertex.  We have actually taken into account the
contributions from diagrams with three effective vertices in our
calculation as well as the dependence of the top decay width on the
effective couplings, but we have explicitly verified in all cases that
the contribution to the cross section from terms of order higher than
$\mathcal{O}(\Lambda^{-4})$ is actually negligible for values of the
effective couplings within the bounds given below.  (We remark here,
parenthetically, that the contributions to the cross section at order
$1/\Lambda^4$ from dimension-eight operators interfering with the SM are
currently unknown and constitute an inherent uncertainty of the EFT
analysis at dimension six.)

\subsection{Methodology and assumptions}
\label{sec:method}

In order to obtain bounds on the effective couplings, we consider the
ratio of the cross section $\sigma_\mathrm{eff}(\{\C_\mathcal{O}\})$
obtained from the Lagrangian (\ref{eq:lag}) at tree level to the SM
cross section $\sigma_\mathrm{SM}=\sigma_\mathrm{eff}(\{0\})$ 
\begin{equation}
  \label{eq:rat}
  R=\frac{\sigma_\mathrm{eff}(\{\C_\mathcal{O}\})}{\sigma_\mathrm{SM}},
\end{equation}
where $\{\C_\mathcal{O}\}$ is the set of anomalous coupling constants.
For a given relative experimental uncertainty $\veps_\mathrm{exp}$,
the region of allowed values for the effective couplings
$\{\C_\mathcal{O}\}$ is determined at the $n_\sigma\;\sigma$
($n_\sigma=1,$ 2,\ldots) level by the inequalities
\begin{equation}
  \label{eq:ineq}
  R \lessgtr 1 \pm n_\sigma\veps_\mathrm{exp}.
\end{equation}
Since only three real effective couplings enter the Lagrangian for the
processes considered here, we can parametrize the ratio
(\ref{eq:rat}) as 
\begin{equation}
  \label{eq:rat0}
  R= 1 + a_1\, \cubr + a_2 \left(\cubr\right)^2 + a'_2
  \left(\cubi\right)^2  + b_1\, \cmpq + b_2
  \left(\cmpq\right)^2 + c\,  \cubr\cmpq + \cdots
\end{equation}
We obtain allowed intervals on the effective couplings by substituting
(\ref{eq:rat0}) in (\ref{eq:ineq}) with only one of the couplings in
(\ref{eq:rat0}) taken to be nonzero. Similarly, we consider also
allowed two-coupling regions for pairs of effective couplings by
setting the remaining one to zero in (\ref{eq:rat0}).  The parameters
in (\ref{eq:rat0}) are determined from an extensive set of Monte Carlo
simulations at the detector level, for each energy and photoproduction
region, to which (\ref{eq:rat0}) is fitted.  Once those parameters are
known, (\ref{eq:ineq}) yields the desired one- or two-dimensional
limits on the effective couplings being considered.  The consistency
condition that the contribution to the cross section from terms of
$O\left(\Lambda^{-6}\right)$ and higher in (\ref{eq:lag}) be
negligibly small entails on the parametrizations (\ref{eq:rat0}) the
requirement that the terms of $O(\C^{3})$ and higher must be
correspondingly negligible within the allowed region determined by
(\ref{eq:ineq}).  We check this consistency condition in all cases
considered below.

The cross-section ratio (\ref{eq:rat}) depends on the
renormalization and factorization scales, for moderate values of the
effective couplings, much more weakly than the cross sections
themselves.  This was previously noticed, in a rather different
context, in \cite{sar14} (see section IV C of that reference).  Thus, the
bounds on the effective couplings obtained from (\ref{eq:ineq}) are
therefore largely independent of those scales.

\begin{table}
\centering{}  
\begin{tabular}{c|ccc|ccc|c}
\multicolumn{1}{c}{} &\multicolumn{3}{c}{\mbox{LHeC}}&\multicolumn{3}{c}{\mbox{FCC-he}}&\\
(\%)&$PhP_{I}$&$PhP_{II}$&$PhP_{III}$&$PhP_{I}$&$PhP_{II}$&$PhP_{III}$&\mbox{section}\\\hline
$tbW$       &10.3&11.3&12.2&10.3&11.4&13.5&\ref{sec:php.mc}\\
$tbq$       & 3.6& 4.4& 4.6& 0.9& 1.2& 1.3&\ref{sec:tbq}\\  
$Whq+WZq$   & 1.8& 1.8& 1.4& 0.2& 0.2& 0.2&\ref{sec:notop.q}\\
$tbW$(red.) &1.0&0.8&1.0&0.2&0.1&0.05&\ref{sec:tbw.red}\\    
statistical&5.0&3.7&2.8&1.5&1.3&1.0&\ref{sec:php.mc}\\\hline
RMS&12.2&12.8&13.4&10.5&11.5&13.6& 
  \end{tabular}
  \caption{Background-to-signal ratios expressed as percentages of the
    signal cross section, for the four main background processes and
    statistical uncertainty.  ``RMS'' refers to the sum in quadrature
    of the previous lines.}
  \label{tab:summary}  
\end{table}
In order to obtain bounds on the effective couplings through
(\ref{eq:ineq}), below we assume $\veps_\mathrm{exp}$ to take values
within a certain interval, which is motivated by estimating the
uncertainties in the signal cross section in our Monte Carlo
simulations through the addition in quadrature of the statistical
uncertainty and the backgrounds cross sections, as summarized in table
\ref{tab:summary}. In view of those results, we will assume total
measurement uncertainties of $\varepsilon_\mathrm{exp}=$12\%, 15\%,
18\%.  The first value would be applicable to photoproduction region
$PhP_{I}$, especially at the FCC-he, the second and third value could
be applicable in the other photoproduction regions.  The largest value
of 18\% is the one used in \cite{bou13b}.  More importantly, once
bounds on the anomalous couplings have been established for these
three uncertainty values, results for other $\veps_\mathrm{exp}$ can
be obtained by interpolation.

\subsection{Bounds on $\boldsymbol{\cmpq}$}
\label{sec:cmpq.results}

We obtain bounds on the left-handed vector $tbW$ coupling from Monte
Carlo simulations as described previously in this section. In table
\ref{tab:bound.cpq} we report the single-coupling bounds for $\cmpq$
obtained from (\ref{eq:ineq}) at the LHeC and FCC-he energies, at the
three photoproduction regions, at the $1\sigma$ level, for three
values of the assumed experimental uncertainty.
\begin{table}
  \centering{}
  \begin{tabular}{cccc|ccc}\cline{2-7}
    \multicolumn{7}{c}{$\cmpq$, 68\% C.L.}\\\cline{2-7}
    & \multicolumn{3}{c}{\mbox{LHeC}}&
                                       \multicolumn{3}{c}{\mbox{FCC-he}}\\\cline{2-7}
    $\varepsilon_\mathrm{exp}$&$PhP_{I}$&$PhP_{II}$&$PhP_{III}$&$PhP_{I}$&$PhP_{II}$&$PhP_{III}$\\\hline
    12\% &-0.11,0.080&-0.056,0.049&-0.039,0.035&-0.11,0.081&-0.061,0.051&-0.039,0.035\\             
    15\% &-0.14,0.098&-0.072,0.060&-0.049,0.043&-0.14,0.099&-0.078,0.063&-0.049,0.043\\             
    18\% &-0.18,0.11 &-0.089,0.071&-0.060,0.052&-0.18,0.12 &-0.097,0.074&-0.060,0.051\\\hline       
  \end{tabular}
  \caption{Single-coupling limits on the charged-current effective
    coupling $\cmpq$, at 68\% C.L., at the LHeC and FCC-he energies,
    in three photoproduction phase-space regions and for three assumed
    experimental uncertainties.}
  \label{tab:bound.cpq}
\end{table}
Clearly, the largest sensitivity is obtained in region $PhP_{III}$.
Indeed, the anomalous coupling $\cmpq$ constitutes a perturbation
$\delta f_V^L$ to the SM CC coupling $f_V^L=1+\delta f_V^L$ and,
therefore, it also perturbs the gauge cancellation discussed above
under eq.\ (\ref{eq:ttx.proc}).  Thus, the sensitivity is largest in
region $PhP_{III}$ where the cancellation is strongest. For that
region we report 95\% C.L.\ results both for the LHeC and the FCC-he
in table \ref{tab:bound.cpq.2}.
\begin{table}
  \centering{}
  \begin{tabular}{ccc}\cline{2-3}
    &  \multicolumn{2}{c}{$\cmpq$, $PhP_{III}$, 95\% C.L.}\\\cline{2-3}
    $\varepsilon_\mathrm{exp}$ & \mbox{LHeC}& \mbox{FCC-he} \\\hline
    12\% &-0.083,0.067&-0.083,0.067\\             
    15\% &-0.11,0.083&-0.11,0.082\\             
    18\% &-0.14,0.098&-0.14,0.097\\\hline       
  \end{tabular}
  \caption{Single-coupling limits on the charged-current effective
    coupling $\cmpq$, at 95\% C.L., at the LHeC and FCC-he energies,
    in photoproduction region $PhP_{III}$ and for three assumed
    experimental uncertainties.}    
  \label{tab:bound.cpq.2}    
\end{table}  
We see from tables \ref{tab:bound.cpq}, \ref{tab:bound.cpq.2} that the
sensitivities obtained at the LHeC and at the FCC-he are essentially
the same. (In appendix \ref{sec:app.bound} we give the bounds for
LHeC, $PhP_{III}$, in terms of $\ccmpq$ for the purpose of comparison
with other bounds in the literature.)  Finally, the fit parameters
from (\ref{eq:rat0}) are given by,
\begin{equation}
  \label{eq:fit.cpq}
  \begin{array}{ccc}
                 &b_1&b_2\\    
    \mbox{LHeC:}&3.259&4.407\\
    \mbox{FCC-he:}&2.263&4.552
  \end{array},
\end{equation}
for the photoproduction region $PhP_{III}$.

A fit similar to (\ref{eq:fit.cpq}) for the main background $tbW$
yields $b_1=2.221$, $b_2=1.650$ at the LHeC, and $b_1=2.275$,
$b_2=1.663$ at the FCC-he. It should be taken into account that, due
to the fact that the cross section for $tbW$ is suppressed by the cuts
(\ref{eq:C0})--(\ref{eq:C4}), these coefficients are subject to a
larger numerical uncertainty than those for the signal in
(\ref{eq:fit.cpq}). Let us consider, for concreteness, the range of
values for $\cmpq$ in table \ref{tab:bound.cpq} corresponding to a
variation $\Delta\sigma_{t\bar{t}}/\sigma_{t\bar{t}}=12\%$ at the LHeC
in region $PhP_{III}$. The parameters $b_{1,2}$ given above for
$\sigma_{tbW}$ imply a corresponding variation
$\Delta\sigma_{tbW}/\sigma_{tbW}=8\%$. Relative to the signal cross
section, see (\ref{eq:sm.xsctn.numeric}), the variation is small
$\Delta\sigma_{tbW}/\sigma_{t\bar{t}}=0.97\%$, for a total uncertainty
of $\sqrt{0.12^2+0.0097^2}=12.04\%$. Similarly small results are found
in the other regions, and at the FCC-he energy.\footnote{This
  paragraph is included here in fulfillment of a referee's
  requirement.}
\label{page.footnote}

It is of interest to compare our results for $\cmpq$ with those
reported by CMS.  A recent measurement of the single-top production
cross section used to set bounds on $\delta f_V^L$ is given in
\cite{cms17a}.  From figure 6 of \cite{cms17a} we obtain the following
limits at 1- and 2-$\sigma$ significance:
\begin{equation}
  \label{eq:cms17a.fVL.new}
\begin{array}{lcc}
     & \mbox{68\% C.L.}& \mbox{95\% C.L.}\\
    \mbox{CMS:} & -0.024 < \delta f_V^L < 0.094 
    & -0.062 < \delta f_V^L < 0.132~,
\end{array}
\end{equation}
with $f_V^L=1+\delta f_V^L$ and, as shown above,
$\delta f_V^L = \cmpq$. We compare the limits we obtain in region
$PhP_{III}$ in tables \ref{tab:bound.cpq} and \ref{tab:bound.cpq.2} to
the CMS result (\ref{eq:cms17a.fVL.new}) by comparing the interval
lengths.  We see that at 68\% C.L.\ our limits are more restrictive
than those from (\ref{eq:cms17a.fVL.new}) for all three values of
$\varepsilon_\mathrm{exp}$ considered in the table.  At
$\varepsilon_\mathrm{exp}=12\%$, in particular, the interval length in
table \ref{tab:bound.cpq} is smaller than that in
(\ref{eq:cms17a.fVL.new}) by roughly one-third.  At 95\% C.L.\ our
results give a tighter bound for $\veps_\mathrm{exp}=12\%$, an equally
tight bound at $\veps_\mathrm{exp}=15\%$ and somewhat looser bounds at
$\veps_\mathrm{exp}=18\%$.

\subsection{Bounds on $\boldsymbol{\cub}$: single-coupling bounds}
\label{sec:cub.results.1}

We obtain bounds on the dipole $tt\gamma$ couplings from Monte Carlo
simulations as previously described in this section. In table
\ref{tab:bound.cubr} we report the single-coupling bounds for $\cubr$
obtained from (\ref{eq:ineq}) at the LHeC and FCC-he energies, in the
three photoproduction regions, at the $1\sigma$ level, for three
values of the assumed experimental uncertainty.
\begin{table} \centering{}
  \begin{tabular}{cccc|ccc}\cline{2-7} \multicolumn{7}{c}{$\cubr$,
68\%\mbox{C.L.}}\\\cline{2-7} & \multicolumn{3}{c}{LHeC}&
\multicolumn{3}{c}{FCC-he}\\\cline{2-7}
$\varepsilon_\mathrm{exp}$&$PhP_{I}$&$PhP_{II}$&$PhP_{III}$&$PhP_{I}$&$PhP_{II}$&$PhP_{III}$\\\hline
12\%&-0.041,0.049&-0.063,0.084&-0.13,0.58&-0.040,0.049&-0.061,0.090&-0.13,0.52\\
15\%&-0.050,0.063&-0.077,0.120&-0.16,0.61&-0.048,0.063&-0.075,0.130&-0.15,0.54\\
18\%&-0.059,0.078&-0.090,0.150&-0.18,0.64&-0.057,0.080&-0.088,0.510&-0.17,0.57\\\hline
  \end{tabular}
  \caption{Single-coupling limits on the effective coupling $\cubr$,
at 68\% C.L., at the LHeC and FCC-he energies, in three
photoproduction phase-space regions and for three assumed experimental
uncertainties.}
  \label{tab:bound.cubr}
\end{table}
Similarly, in table \ref{tab:bound.cubi} we show the
bounds for the coupling $\widetilde{C}_{uB\,i}^{33}$.
  \begin{table} \centering{}
    \begin{tabular}{cccc|ccc}\cline{2-7} \multicolumn{7}{c}{$\cubi$,
68\%\mbox{C.L.}}\\\cline{2-7} & \multicolumn{3}{c|}{LHeC}&
\multicolumn{3}{c}{FCC-he}\\\cline{2-7}
$\varepsilon_\mathrm{exp}$&$PhP_{I}$&$PhP_{II}$&$PhP_{III}$&$PhP_{I}$&$PhP_{II}$&$PhP_{III}$\\\hline
12\% &$\pm$0.15&$\pm$0.18&$\pm$0.28&$\pm$0.13&$\pm$0.18&$\pm$0.25\\
15\% &$\pm$0.16&$\pm$0.21&$\pm$0.32&$\pm$0.15&$\pm$0.20&$\pm$0.28\\
18\%
&$\pm$0.18&$\pm$0.23&$\pm$0.35&$\pm$0.17&$\pm$0.22&$\pm$0.31\\\hline
    \end{tabular}
  \caption{Single-coupling limits on the effective coupling $\cubi$,
at 68\% C.L., at the LHeC and FCC-he energies, in three
photoproduction phase-space regions and for three assumed experimental
uncertainties.}
  \label{tab:bound.cubi}
\end{table} Clearly, the largest sensitivity to
$\widetilde{C}_{uB}^{33}$ is obtained in region $PhP_{I}$. This is due
to two different mechanisms.  First, the fact that the SM is close to
an infrared divergence at $Q^2=0$ and, therefore, as $Q^2$ decreases
the SM cross section grows much faster than the dipolar cross section,
which is infrared finite.  This causes the sensitivity to both $\cubr$,
$\cubi$ to decrease as we go from $PhP_{I}$ to $PhP_{III}$.  Second,
as seen in figure \ref{fig:escat} (see left panel), the rapidity
distribution of the scattered electron is more sharply peaked in
$PhP_{III}$ than in $PhP_{I}$, thus leading to a smaller interference
with the dipolar amplitude that leads to a flat distribution for
$y(e^-)$.  This causes a reduction in the sensitivity to $\cubr$ in
$PhP_{III}$ as compared to $PhP_{I}$.

The results obtained in region $PhP_{I}$ are also
given at the 2$\sigma$ level in table \ref{tab:bound.cubr.2}. (In
appendix \ref{sec:app.bound} we give the bounds for LHeC, $PhP_{I}$,
in terms of $\ccubr$ for the purpose of comparison with other bounds
in the literature.)
\begin{table} \centering{}
  \begin{tabular}{ccc|cc}\cline{2-5}
&\multicolumn{4}{c}{$PhP_{I}$,95\%\mbox{C.L.}}\\\cline{2-5} & \multicolumn{2}{c|}{$\cubr$}
                                                            &\multicolumn{2}{c}{$\cubi$}\\\cline{2-5}
$\varepsilon_\mathrm{exp}$ & LHeC & FCC-he & LHeC& FCC-he \\\hline
    12\% &-0.076,0.11&-0.074,0.12&$\pm$0.21&$\pm$0.19\\
    15\% &-0.092,0.17&-0.088,0.21&$\pm$0.23&$\pm$0.21\\
    18\% &-0.110,0.60&-0.100,0.53&$\pm$0.25&$\pm$0.23\\\hline
  \end{tabular}
  \caption{Single-coupling limits on the effective couplings $\cubr$,
$\cubi$, at 95\% C.L., at the LHeC and FCC-he energies, in
photoproduction region $PhP_{I}$ and for three assumed experimental
uncertainties.}
  \label{tab:bound.cubr.2}
  \end{table}
Finally, the fit parameters from (\ref{eq:rat0}) are given by,
\begin{equation}
  \label{eq:fit.cub}
  \begin{array}{cccc}
                  &a_1&a_2   & a'_2\\    
    \mbox{LHeC:}  &-2.733&5.547&5.585\\    
    \mbox{FCC-he:}&-2.778&6.509&6.586
  \end{array},
\end{equation}
for the photoproduction region $PhP_{I}$.

In photoproduction region $PhP_I$ the results analogous to
(\ref{eq:fit.cub}) for the main background process $tbW$ are given by
$a_1=-0.177$, $a_2=0.0599$, $a'_2=0.198$ at the LHeC and $a_1=-0.135$,
$a_2=0.271$, $a'_2=0.364$ at the FCC-he. These numbers are subject to
a larger numerical uncertainty than those for the signal in
(\ref{eq:fit.cub}), because the cross section for $tbW$ is suppressed
by the cuts (\ref{eq:C0})--(\ref{eq:C4}).  In this case, the range of
values for $\cubr$ in table \ref{tab:bound.cubr} corresponding to a
variation $\Delta\sigma_{t\bar{t}}/\sigma_{t\bar{t}}=12\%$ at the LHeC
in region $PhP_{I}$, leads to $\Delta\sigma_{tbW}/\sigma_{tbW}=0.8\%$
which, relative to the signal cross section, corresponds to
$\Delta\sigma_{tbW}/\sigma_{t\bar{t}}=0.08\%$.  Similarly negligibly
small results are found in the other regions, and at the FCC-he
energy.\footnote{See footnote on page \pageref{page.footnote}.}
  
We mention that although not reported in detail here, we have carried
out a complete parton-level analysis of the sensitivity to $\cubr$,
$\cubi$ in all three photoproduction regions.  In the case of region
$PhP_{I}$, the limits obtained on those couplings are essentially the
same as those obtained previously in \cite{bou13b} in the framework of
the EPA. On the other hand, it is of interest to compare the bounds
obtained here from parametric detector-level Monte Carlo simulations
to those from the parton-level analysis of \cite{bou13b}. From table
IX of \cite{bou13b}, and using the relation (\ref{eq:relations})
between $\kappa$, $\widetilde{\kappa}$ and $\cubr$, $\cubi$, we find
that, assuming $\varepsilon_\mathrm{exp}=10$\% experimental uncertainty and at
1$\sigma$ C.L., we have,
\begin{equation}
  \label{eq:boubou5}
 -0.022  <  \widetilde{C}_{uB\,r}^{33} < 0.026,
\quad
 |\widetilde{C}_{uB\,i}^{33}| < 0.10.
\end{equation}
From the same table in \cite{bou13b} we find the following bounds also
at 1$\sigma$ and at $\varepsilon=18$\% experimental uncertainty,
\begin{equation}
  \label{eq:boubou7}
 -0.039  <  \widetilde{C}_{uB\,r}^{33} < 0.047,
\quad
 |\widetilde{C}_{uB\,i}^{33}| < 0.14.
\end{equation}
These bounds are tighter than the comparable ones in tables
\ref{tab:bound.cubr}, \ref{tab:bound.cubi}, as expected since the
analysis in \cite{bou13b} is carried out at the parton level only.  We
notice, however, that the more complete analysis performed here yields
68\% C.L.\ intervals for $\widetilde{C}_{uB\,r}^{33}$ that are
slightly less than 50\% larger, and for $\widetilde{C}_{uB\,i}^{33}$
only 33\% larger, than the partonic analysis in the EPA in
\cite{bou13b}.

\subsection{Two-dimensional allowed regions}
\label{sec:cub.results.2}

\begin{figure}[!t]
  \centering{}
\begin{picture}(450,400)(0,0)
\put(0,0){\includegraphics[scale=0.95]{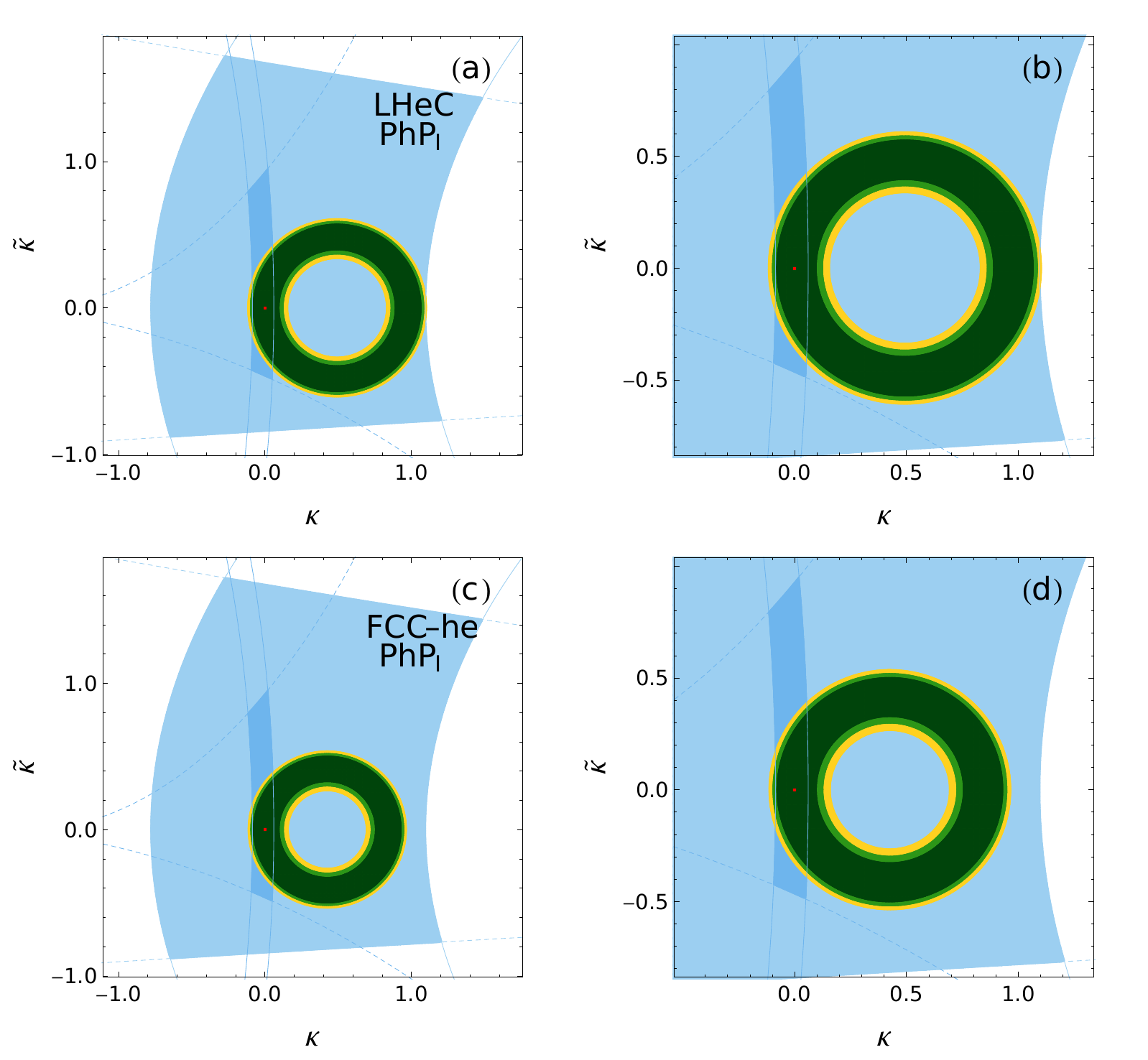}}
\end{picture}
\caption{Allowed regions for the
      top quark dipole moments $\kappa$ and $\widetilde{\kappa}$ at
      (a),(b) the LHeC and (c),(d) the FCC-he. Panels (a), (c) display
      a global view, (b), (d) a magnified one.  Annular regions:
      regions allowed at 68\% C.L.\ by a top-pair
      tagged-photoproduction cross-section measurement, in
      photoproduction region $PhP_{I}$ (\ref{eq:php}), with
      experimental uncertainties 12\% (dark green), 15\% (light
      green), and 18\% (yellow). Light-blue
      area: region allowed by the measurements of the branching ratio
      and $CP$ asymmetry of $B\rightarrow X_s\gamma$ decays, from
      inequalities (\ref{bsgregion}), (\ref{asymregion}) with
      (\ref{br.hewett.top}), (\ref{asyminkappas.1top}).  Darker-blue
      area: same as previous, but with (\ref{br.crivel.top}),
      (\ref{asyminkappas.2top}).}
\label{fig:1}
\end{figure}

In figure \ref{fig:1} we show the allowed regions in the
$\kappa$--$\widetilde{\kappa}$ plane, determined by the top-pair
photoproduction cross section at both the LHeC and FCC-he energies, in
region $PhP_{I}$ at 68\% C.L.  These regions are given by the circular
coronas determined by (\ref{eq:ineq}) with the parametrization
(\ref{eq:rat0}) with the parameters (\ref{eq:fit.cub}).  The allowed
regions shown in figure \ref{fig:1} correspond to the assumed
measurement uncertainties $\veps_\mathrm{exp}=12$, 15, 18\% in
different colors as indicated in the figure caption.  Also shown in
the figure are the regions in the $\kappa$--$\widetilde{\kappa}$ plane
allowed by the branching ratio and $CP$ asymmetry for the process
$B\rightarrow X_s\gamma$ through inequalities (\ref{bsgregion}),
(\ref{asymregion}) in both the form obtained from \cite{hewett} as
given in (\ref{br.hewett.top}), (\ref{asyminkappas.1top}), and the
form from \cite{crivellin15} given by (\ref{br.crivel.top}),
(\ref{asyminkappas.2top}). The difference in area between these two
regions hardly needs to be emphasized. We remark, however, that even
the smaller region resulting from (\ref{br.crivel.top}),
(\ref{asyminkappas.2top}) is not completely contained in the annular
regions determined by top-pair photoproduction, which results in a
significant reduction of the allowed parameter space.

Also seen in figure \ref{fig:1} is that the annular allowed regions
obtained at the FCC-he are somewhat smaller than those at the LHeC
energy.  We notice, however, that both sets of allowed regions are
identical in the neighborhood of the origin (i.e., the SM), which is
consistent with the individual-coupling bounds shown in tables
\ref{tab:bound.cubr}, \ref{tab:bound.cubi}, \ref{tab:bound.cubr.2}
being the same at both energies.  We can also make a comparison of
figure \ref{fig:1} with figure 5 of \cite{bou13b} (see also figure 71
of \cite{LHeC2020}), obtained at the parton level and in the EPA.  The
region allowed by ${\bar B} \to X_s \gamma$ in figure 5 of
\cite{bou13b} covers a region of approximately $-2.4 < \kappa < 0.3$
largely shifted towards negative values.  That was because
$10^4 {\rm BR}^{\rm exp}{\bar B} \to X_s \gamma = 3.43$ was about
$10\%$ larger than
$10^4 {\rm BR}^{\rm theo}{\bar B} \to X_s \gamma = 3.15$ and then
large negative values of $\kappa$ would increase the theoretical value
and bring it closer to the experiment (see eq.\ (\ref{br.hewett})).
In figure \ref{fig:1} the situation has reversed and now
$10^4 {\rm BR}^{\rm exp}{\bar B} \to X_s \gamma = 3.32$ is about $1\%$
smaller than $10^4 {\rm BR}^{\rm theo}{\bar B} \to X_s \gamma = 3.36$
and then $\kappa$ tends to lean somewhat toward positive values.  On
the other hand, the $18\%$ error region allowed by top-pair
photoproduction in figure 5 \cite{bou13b} presents the shape of a
corona roughly $0.1$ in thickness, whereas now in figure \ref{fig:1}
the thickness is about $0.15$ for the same $18\%$ experimental
uncertainty. This loss of sensitivity is due to the transition from
parton-level to detector-level simulations, and was to be expected.

\begin{figure}[th]
  \centering{}
\begin{picture}(450,200)(0,0)
\put(0,0){\includegraphics[scale=1.0]{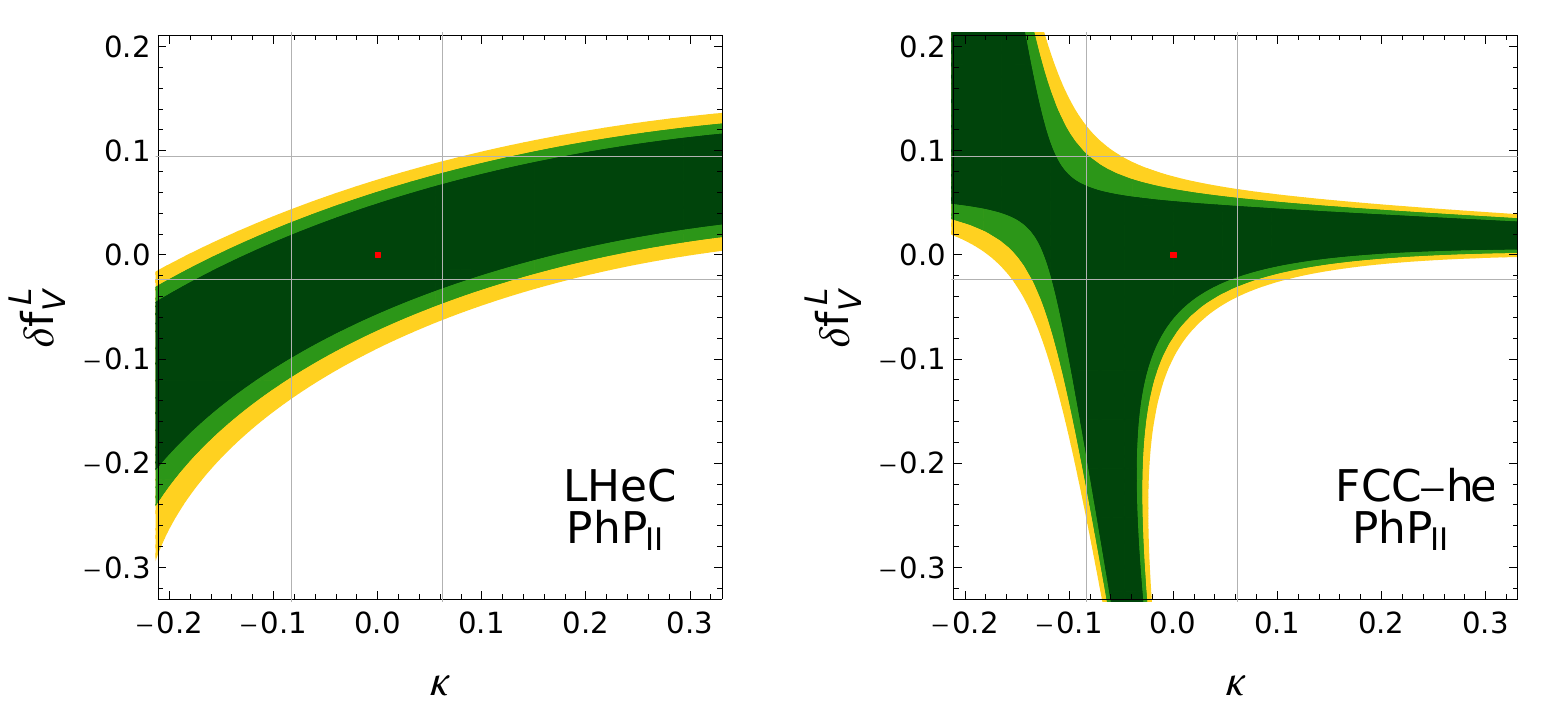}}
\end{picture}
\caption{Allowed regions for the
      top quark dipole moment $\kappa$ and the CC form-factor
      $\delta f_{V}^{L}$ at the LHeC (left panel) and FCC-he (right
      panel) in photoproduction region $PhP_{II}$
      (\ref{eq:php}), at 68\% C.L.\ by a top-pair 
      tagged-photoproduction cross-section measurement with
      experimental uncertainties 12\% (dark green), 15\% (light
      green), and 18\% (yellow).  Vertical
      lines: single-coupling limits for $\kappa$ from the branching
      ratio for $B\rightarrow X_s\gamma$ decays,
      (\ref{kappatlimits.2}). Horizontal lines: single-coupling limits
      for $\delta f_{V}^{L}$ from (\ref{eq:cms17a.fVL.new})
      \cite{cms17a}.  }
\label{fig:1a}
\end{figure}

It is also desirable to obtain the allowed regions in the
$\cubr$--$\cmpq$ plane or, through (\ref{eq:relations}) the 
$\kappa$--$\delta f_V^L$ plane.  We notice here, however, that in
photoproduction region $PhP_{I}$, where the sensitivity to $\kappa$ is
highest, the corresponding sensitivity to $\delta f_V^L$ is too
low to determine a closed region in the plane.  Similarly, in
photoproduction region $PhP_{III}$, the sensitivity to $\delta f_V^L$ is
highest, but the sensitivity to $\kappa$ is too low to yield a closed
region. We therefore use the region $PhP_{II}$, where the
sensitivities to those couplings are not optimal for each single
coupling, but high enough for both to obtain a closed region.  The
parametrization (\ref{eq:rat0}) holds in the kinematic region
$PhP_{II}$ with parameters,
\begin{equation}
  \label{eq:fit.II}
  \begin{array}{cccccc}
                  &a_1   &a_2   &b_1  &b_2   &c\\    
    \mbox{LHeC:}  &-1.706&3.319&2.313&3.221&0.0218\\
    \mbox{FCC-he:}&-1.693&4.018&2.183&3.399&0.155 
  \end{array}.
\end{equation}
The parameters in this equation comparable to those in
(\ref{eq:fit.cpq}), (\ref{eq:fit.cub}) are seen to be significantly
smaller, reflecting the reduced sensitivity in this region to both
couplings. The allowed region in the $\kappa$--$\delta f_V^L$ plane obtained from
(\ref{eq:fit.II}) with $\varepsilon_\mathrm{exp}=12\%$ at 68\% C.L.\
is shown in figure \ref{fig:1a}. As seen there, the sensitivities to
either $\cubr$ or $\cmpq$ are not strong enough to properly constrain
the two-dimensional parameter space. In the case of the LHeC, the
interference coefficient $c$ in (\ref{eq:fit.II}) is small enough that
the regions determined by each inequality in (\ref{eq:ineq}) are
ellipses, whose intersection leads to the large annular allowed region
shown in the left panel of fig.\ \ref{fig:1a}. At the FCC-he energy
the parameter $c$ in (\ref{eq:fit.II}) is comparatively much larger, so
the regions determined by (\ref{eq:ineq}) are now hyperbolas,
whose intersection is the star-shaped allowed region seen in the right
panel of fig.\ \ref{fig:1a}.
Also shown in the figure are the
individual-coupling limits for $\delta f_V^L$ from
(\ref{eq:cms17a.fVL.new}) \cite{cms17a}, and for $\kappa$ from
(\ref{kappatlimits.2}).  We see that even with the reduced sensitivity
in region $PhP_{II}$, the allowed region determined by the top-pair
photoproduction process cuts part of the square region determined by
the individual-coupling limits.

\section{Final remarks}
\label{sec:final}

In this paper we presented a dedicated study of top-pair photoproduction
in semileptonic mode in $pe^-$ collisions at the LHeC and FCC-he
future colliders.  We performed an extensive set of Monte Carlo
simulations at the fast detector simulation level of that process and
its SM backgrounds.  The most relevant background processes are found
to be the irreducible $tbW$, $tbq$, and $Whq$ photoproduction.

In our cross-section computations all resonant and nonresonant
diagrams are taken into account, and all off-shell effects for the top
quark and $W$ boson, and $Z$ and $h$ bosons in the background
processes, are included. Furthermore, since we perform our
calculations with the full QED scattering amplitude (not relying on
the EPA) we take into account the complete photoproduction kinematics.
This allows us to define three photoproduction regions based on the
angular acceptance range of the electron tagger.  We find that those
regions provide different sensitivity levels to different top quark
effective couplings.  Another consequence of adopting the framework of
full tree-level QED is that, besides the electromagnetic dipole
$t\bar{t}\gamma$ coupling that is our main interest, we find
significant sensitivity to the SM-like left-handed vector $tbW$
coupling, for which photoproduction also turns out to be a good probe.

We find in section \ref{sec:res.coupl} that the sensitivity of
top-pair photoproduction at the LHeC and FCC-he to the top e.m.\
dipole moments is highest in photoproduction region $PhP_{I}$ (see
(\ref{eq:php})), moderate in $PhP_{II}$ and poor in $PhP_{III}$. The
mechanisms causing this are discussed in section
\ref{sec:cub.results.1}.  An analysis of this process in $PhP_{I}$ at
the parton level yields essentially the same results as obtained
previously in \cite{bou13b}.  The more realistic simulations carried
out here lead, of course, to somewhat weaker limits. Our results are
therefore consistent with those of \cite{bou13b}. However, the
phase-space region of validity of the results of \cite{bou13b}, region
$PhP_{I}$, could not have been determined with the methods used there.

Another important result (that could not have been obtained in
\cite{bou13b}) is that top-pair photoproduction at the LHeC and FCC-he
has significant sensitivity to the $tbW$ effective coupling
$\cmpq=\delta f_V^L$ (alternatively, $\ccmpq$). The limits obtained on
this coupling are strongest in region $PhP_{III}$, moderate in
$PhP_{II}$ and poor in $PhP_{I}$.  The reasons why this is so are
discussed in section \ref{sec:cmpq.results}, and they directly imply
that a higher sensitivity could be obtained if we could attain angular
acceptances down to angles smaller than the 4 mrad from the beam
assumed in $PhP_{III}$. While this is true, we point out here that
such sensitivity gains will encounter diminishing returns as the
minimal scattering angle (measured from the $e^-$-beam direction) is
decreased from its value in $PhP_{III}$.

It is apparent from section \ref{sec:res.coupl} that the strength of
the dependence of the top-pair photoproduction cross section on the
effective couplings is essentially the same at the LHeC and the
FCC-he. This leads to the same sensitivity at both colliders if we
assume the same measurement uncertainty for both, as seen in the
individual-coupling limits reported in section \ref{sec:res.coupl}. We
notice, however, that the statistical uncertainties are smaller at the
FCC-he than at the LHeC, due to the larger cross sections (see
(\ref{eq:sm.xsctn.numeric})), which is important especially in
photoproduction region $PhP_{I}$, where the systematic uncertainty from
backgrounds is also slightly smaller at the FCC-he (see table
\ref{tab:summary}). Thus, we may expect a somewhat smaller uncertainty
and a slightly larger sensitivity to the top quark e.m.\ dipole
moments at the FCC-he than at the LHeC.  In the other two
photoproduction regions the effects of the statistical uncertainty are
relatively less important, so we expect the same sensitivity from both
colliders.

In section \ref{sec:overview} we summarize the results of five
independent current global analyses of top quark effective couplings
that have been published in the last two years (see table
\ref{tab:global}).  We find there that two of those studies report
very weak constraints on $C^{33}_{uB}$.  Another two do not report
limits for this coefficient at all; and then, one of them reports
significantly stronger constraints, and the reason is that it is the
only one that includes indirect limits from the
${\bar B} \to X_s \gamma$ branching ratio.

From a quantitative point of view, the main results of this paper are
the SM cross sections obtained in section \ref{sec:php.mc} (see eq.\
(\ref{eq:sm.xsctn.numeric})) for the $t\bar{t}$ and $tbW$
photoproduction processes, at the LHeC and FCC-he and in the three
photoproduction regions, as well as the various differential cross
sections displayed in the figures in that section. The extensive
analysis of backgrounds performed in section \ref{sec:sm.bck} is
summarized in table \ref{tab:summary}, which is also a relevant
result. The most important results, however, are the
individual-coupling limits given in section \ref{sec:cmpq.results}
(see tables \ref{tab:bound.cpq}, \ref{tab:bound.cpq.2}) and section
\ref{sec:cub.results.1} (see tables \ref{tab:bound.cubr},
\ref{tab:bound.cubi}, \ref{tab:bound.cubr.2}), also given at two
energies and three photoproduction regions, as well as the limits in
section \ref{sec:branch} (see eqs.\ (\ref{kappatlimits.1}),
(\ref{kappatlimits.2})), section \ref{sec:asym} (see eqs.\
(\ref{asymzz.1}), (\ref{asymzz.2})), and the two-dimensional allowed
regions found in section \ref{sec:cub.results.2} (see figures
\ref{fig:1}, \ref{fig:1a}).

Taken together, these results show that measurements of top-pair
photoproduction at the LHeC and FCC-he will lead to tight direct
bounds on top quark e.m.\ dipole moments, greatly improving on the
direct limits resulting from hadron-hadron colliders present and
future.  As for the $tbW$ left-handed vector coupling $\ccmpq$ (or
$\delta f_V^L$), our results strongly suggest that the LHeC and FCC-he
measurement will result in limits tighter than the current ones from
the LHC, and probably as good as those obtained at the HL-LHC.  We
conclude from these observations that measurements of top-pair
photoproduction cross section at the LHeC and FCC-he will provide
greatly valuable contributions to future global analyses.

\paragraph*{Acknowledgments}

This work was partially supported by Sistema Nacional de
Investigadores de Mexico.

\appendix{}

\section{Neutrino momentum reconstruction}
\label{sec:neutr.reco}

In processes like $t\bar{t}$ or $tbW$ photoproduction,
(\ref{eq:ttx.proc}), (\ref{eq:tbw.proc}), in semileptonic mode, there
is a single neutrino in the final state and therefore its transverse
momentum is observable,
$\vec{p}^\perp_\nu=\vec{p}^\perp_\mathrm{miss}$.  We can then use the
kinematics of the decay $W\rightarrow\ell\nu_\ell$ to reconstruct
$p^z_\nu$, assuming that $p_W^2$ is not too far from $m^2_W$.  
We define the $W$-boson transverse mass in the standard way 
(see eq.\ (48.50) in \cite{PDG}),
\begin{equation}
  \label{eq:kin.1}
    p^2_{W\perp} = (|\vec{p}^\perp_\ell|+|\vec{p}^\perp_\nu|)^2-
    (\vec{p}^\perp_\ell+\vec{p}^\perp_\nu)^2 =2 |\vec{p}^\perp_\ell|
    |\vec{p}^\perp_\nu| -
    2 \vec{p}^\perp_\ell\cdot \vec{p}^\perp_\nu, 
\end{equation}
which is an observable quantity. In the massless approximation
we have $p_W^2=2p_\ell\cdot p_\nu$, which together with
(\ref{eq:kin.1}) leads to the relation,
\begin{equation}
  \label{eq:kin.4}
  \frac{1}{2} (p^2_W-p^2_{W\perp}) = |\vec{p}_\ell|
  \sqrt{{p^z_\nu}^2+|\vec{p}^\perp_\nu|^2} - p^z_\ell p^z_\nu -
  |\vec{p}^\perp_\ell| |\vec{p}^\perp_\nu|,
\end{equation}
Squaring both sides of this equality and rearranging terms yields the
quadratic equation
\begin{equation}
  \label{eq:kin.6}
  |\vec{p}^\perp_\ell|^2 {p^z_\nu}^2 - p^z_\ell X^2 p^z_\nu
  +|\vec{p}_\ell|^2 |\vec{p}^\perp_\nu|^2 - \frac{1}{4} X^4 =
  0~,
  \qquad
  \text{with}\;
  X^2 =2|\vec{p}_\ell| |\vec{p}_\nu| + (p^2_W-p^2_{W\perp}),
\end{equation}
which can be solved for $p_\nu^z$. Those solutions can be written in
the form, 
\begin{equation}
  \label{eq:kin.9}
  p^z_{\nu\pm} = \frac{p^z_\ell}{|\vec{p}^\perp_\ell|^2} \left(
    \frac{1}{2} (p^2_{W}-p^2_{W\perp}) + |\vec{p}^\perp_\ell||\vec{p}^\perp_\nu|
  \right)
  \pm
  \frac{|\vec{p}_\ell|}{|\vec{p}^\perp_\ell|^2} \sqrt{p^2_{W}-p^2_{W\perp}}
  \sqrt{\frac{1}{4}(p^2_{W}-p^2_{W\perp}) + |\vec{p}^\perp_\ell| |\vec{p}^\perp_\nu|}~.
\end{equation}
In order to find a numerical value for $p_\nu^z$ we need to assign a
value to $p^2_W$ and to choose one of the two roots in
(\ref{eq:kin.9}).  We address both issues in turn in what follows.

All quantities entering the right-hand side of (\ref{eq:kin.9}) are
observable, with the exception of $p^2_W$.  If we substitute $p^2_W$
by the measured mass ${m^\mathrm{exp}_W}^2$, as is customary in the
literature, the argument of the first radical in (\ref{eq:kin.9})
becomes ${m^\mathrm{exp}_{W}}^2-p^2_{W\perp}$, which is not
necessarily positive.  This results in an implicit cut on events in
which ${m^\mathrm{exp}_{W}}^2<p^2_{W\perp}$.  We choose to make our
cuts fully explicit, so we must choose $p^2_W$ in such a way that
$p^2_{W}-p^2_{W\perp}\geq0$ for all events.  One such possible choice
is
\begin{equation}
  \label{eq:kin.b}
  p^2_W = \max\left\{{{m^\mathrm{exp}_{W}}^2,p^2_{W\perp}} \right\}. 
\end{equation}
We remark, however, that this is the value to be used as a parameter
in (\ref{eq:kin.9}).  The actual value of $p^2_W$ in the event is then
given by the relation $p^2_W=2p_\ell\cdot p_\nu$ with the
reconstructed value of $p^z_\nu$.

Regarding the choice of root in (\ref{eq:kin.9}), we notice that it is
a standard heuristic procedure in the literature to choose $p^z_{\nu}$
as the root with the smaller absolute value.  From (\ref{eq:kin.9}),
that choice is
\begin{equation}
  \label{eq:kin.a}
  p^z_\nu = \left\{
    \begin{array}{cl}
      p^z_{\nu-}& \mathrm{if} \; p^z_\ell >0\\
      p^z_{\nu+}& \mathrm{if} \; p^z_\ell <0
    \end{array}
    \right..
\end{equation}
As this equation should make apparent, the relations
$|p^z_{\nu+}|\lessgtr|p^z_{\nu-}|$ are not invariant under
longitudinal Lorentz boosts.  Thus, the prescription of the root with
``the smaller absolute value'' is valid only in some frames but not in
others.  Explicit computation shows, in fact, that in the lab frame
the correct value of $p^z_{\nu}$ corresponds to the root with the
smaller absolute value in roughly half the events, and to the other
root in the remaining ones. We have found, on the other hand, that the
``smaller absolute value'' prescription works well in the average rest
frame of the top quark.  By this we mean a frame in which the rapidity
distribution of the leptonically decaying top quark is peaked at
$y=0$.  Explicitly, this corresponds to a Lorentz boost in the forward
direction with parameter $\beta=\cosh(1.7)$ at the LHeC (see the
central panel of figure \ref{fig:c9}) and $\beta=\cosh(2.0)$ at the
FCC-he. We stress here, however, that a relation of the form
$|p^z_{\nu+}|>|p^z_{\nu-}|$ defines an open set in the manifold of
proper orthochronous Lorentz transformations, so we do not need to
find a precise value for $\beta$ but, rather, one in the correct
neighborhood.

\begin{figure}
  \centering{}
\includegraphics[scale=1.07]{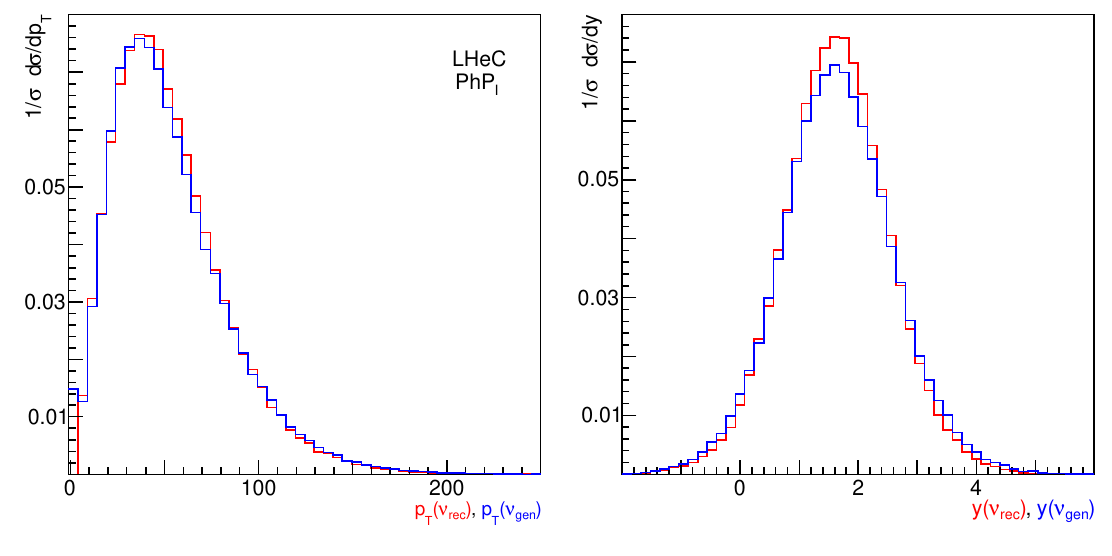}  
  \caption{Normalized differential cross sections with respect to
    transverse momentum and rapidity for the reconstructed neutrino
    (red line) and the generated one (blue line).}
  \label{fig:figapp}
\end{figure}
We assess the goodness of our approach to neutrino reconstruction by
comparing the transverse momentum and the rapidity of the generated
neutrino and the reconstructed one. Here the generated neutrino is
identified by requiring it to be a decay product of a $W$ boson which,
in turn, is a decay product of a top quark. Notice that this is not a
full validation, since the two distributions need not be identical,
but we expect them to be close to each other.  The result of this
comparison is shown in figure \ref{fig:figapp}.  The
transverse-momentum distributions are seen in the figure to be
essentially equal, as expected, and the rapidity distributions are
very close to each other, with the reconstructed neutrino distribution
being slightly narrower, and therefore slightly taller at the maximum.

\section{Allowed ($\kappa$,$\widetilde{\kappa}$) region from
  ${\bar B} \to X_s \gamma$ branching ratio and $CP$ asymmetry.}
\label{sec:b.sa}

In this appendix we obtain bounds on $\kappa$ and $\widetilde{\kappa}$
from a comparison of the experimental and theoretical values of the
${\bar B} \to X_s \gamma$ branching ratio and $CP$ asymmetry.

\subsection{Scale dependence of $C^{33}_{uB}$ and $\kappa$,
  $\tilde \kappa$.}
\label{sec:RG.run.B}

Before entering into the discussion of
${\rm BR} ({\bar B} \to X_s \gamma)$ let us consider
the scale dependence of the effective coefficients.
The contribution of $\kappa$ (and $\tilde \kappa$)
to the effective coefficient $C_7$ is taken (as usual)
at the scale $m_W$  ($\kappa (m_W)$ or $C^{33}_{uB} (m_W)$)
and this is denoted by $\Delta C_7 (m_W)$.
On the other hand, the limits from $t\bar t$ photoproduction
are to be considered at the $m_t$ scale
($\kappa (m_t)$ or $C^{33}_{uB} (m_t)$).  These two scales
are not too far apart and the running produces only a few
percent change in numerical value.
The RGE for the running of $C^{33}_{uB}$, (and indeed also for
all of the other dimension-six effective operators) has
been provided in References
\cite{manohar10} and \cite{manohar12}:
\begin{eqnarray}
  16 \pi^2 \mu \frac{d}{d\mu} C^{33}_{uB} &=&
\left(  \frac{15}{2} y^2_t + \frac{8}{3} g^2_s
-\frac{9}{4} g^2 + (8+25/36) {g'}^2  \right) C^{33}_{uB}
\; , \nonumber \\
{\rm so} \; {\rm that} \;\;\;\; \mu \frac{d}{d\mu} C^{33}_{uB} 
   &\simeq& 0.073 C^{33}_{uB} 
   \; , \nonumber \\
{\rm and}\; {\rm then} \;\;\;\;
C^{33}_{uB} (m_t) &\simeq& 1.06 \; C^{33}_{uB} (m_W) \; ,
\label{rgecub}
\end{eqnarray}
where in the above equations some approximations have
been made.  The only Yukawa factor considered is the
top's $y_t =1$, and the values of the gauge couplings
have been taken as nearly constant and at the scale $m_Z$.
Also, in general there are contributions from other
dimension-six operators, but we are not considering them.
In the context of a global analysis with several
effective operators and a variety of observables, keeping
track of the scale dependence is important because mixing
effects can be significant.
However we do not expect, even in a general context, that
there will be a substantial variation of $C^{33}_{uB}$ in
going from the scale $m_t$ down to $m_W$.
A variation of a few percent given by the factor $1.06$
is small but we will take it into account.

\subsection{The branching ratio $BR ({\bar B} \to X_s \gamma)$}
\label{sec:branch}

So now let us turn our attention to
${\bar B} \to X_s \gamma$.
The ${\bar B} \to X_s \gamma$ branching ratio and
associated $CP$ asymmetry are known to be very good indirect
tests of the anomalous electromagnetic dipole moments of
the top quark as well as many other NP scenarios
(See for instance:
\cite{bou13a}, \cite{bou13b}, \cite{cirigliano16}, 
\cite{moretti}, \cite{crivellin15}).
A recent study that uses $BR ({\bar B} \to X_s \gamma)$
can be found in \cite{bissmannbsg}, and we will use
the same average measurement by \cite{bsgexp} that they used:
\begin{eqnarray}
10^4 BR ({\bar B} \to X_s \gamma)^{\rm exp}_{E\gamma > 1.6 {\rm GeV}} =
3.32 \pm 0.15 \; ,
\end{eqnarray}
and the same NNLO SM calculation
\cite{misiak15} (also \cite{misiak20,czakon15}):
\begin{eqnarray}
10^4 BR ({\bar B} \to X_s \gamma)^{\rm SM}_{E\gamma > 1.6 {\rm GeV}} =
3.36 \pm 0.23 \; .
\end{eqnarray}
The allowed ($\kappa , {\widetilde{\kappa}}$) parameter regions
obtained below are based on these two values.

The ${\bar B} \to X_s \gamma$
branching fraction is given by \cite{misiakbsg}:
\begin{eqnarray}
 BR ({\bar B} \to X_s \gamma)^{\rm th}_{E\gamma > E_0} =
 BR ({\bar B} \to X_c e\bar \nu)^{}_{\rm exp}
 \frac{|V^*_{ts}V_{tb}|}{|V_{cb}|} \frac{6 \alpha_e}{\pi C}
\left( P(E_0) + N(E_0) \right) \; ,
\label{bsgbranching}
\end{eqnarray}
where $E_0=1.6$GeV is the minimum photon energy, $P(E_0)$ and
$N(E_0)$ are the perturbative and nonperturbative contributions.
The constant $C$ is a ratio of ${\bar B} \to X_c e\bar \nu$ and
${\bar B} \to X_u e\bar \nu$ amplitudes times CKM parameters and
it has an experimental value of $0.568$ \cite{alberti}.
The perturbative term is a polynomial in the effective Wilson
coefficients of the effective Hamiltonian at the scale
$\mu_b$ \cite{bissmannbsg,misiakbsg}:
\begin{eqnarray}
  P(E_0) = \sum^{8}_{i,j=1} C_{i}^{\rm eff}(\mu_b)
  C_{j}^{\rm eff}(\mu_b) K_{ij}^{}(E_0,\mu_b) \; ,
\label{pecero}
\end{eqnarray}
with the Wilson coefficients expanded as:
\begin{eqnarray}
 C_{i}^{\rm eff}(\mu_b) =
 C_{j}^{(0)}(\mu_b) +
 \frac{\alpha_s (\mu_b)}{4\pi} C_{j}^{(1)}(\mu_b) +
 \left( \frac{\alpha_s (\mu_b)}{4\pi} \right)^2
 C_{j}^{(2)}(\mu_b) + \cdots~,
\label{ceff}
\end{eqnarray}
where we have omitted electromagnetic correction terms \cite{kagan}.
In particular, for the coefficient $C_7^{(0)}$ \cite{kagan}:
\begin{eqnarray}
  C_{7}^{(0)}(\mu_b) = \eta^{16/23} C_{7}^{(0)}(m_W) +
  \frac{8}{3} (\eta^{14/23}-\eta^{16/23})
 C_{8}^{(0)}(m_W) + \cdots~,
\label{ceffw}
\end{eqnarray}
where $C^{(0)}_{7} (m_W)$ is the LO contribution that comes from the
dimension-four SM Lagrangian as well as the dimension-six operators
of the SMEFT.

We will not attempt to use the lengthy expresions that arise when
using eqs.\ (\ref{bsgbranching})--(\ref{ceffw}).  Instead, we will use
a simplified expresion that was given many years ago in
Ref.~\cite{kagan} (specifically, their eq.\ (20)).  Let us separate
SM and NP contributions $C_7 = C^{\rm SM}_7 + \Delta C^{NP}_7$; then,
we have
\begin{eqnarray}
10^4  BR ({\bar B} \to X_s \gamma)^{\rm th}_{E\gamma > E_0} &=&
10^4 BR ({\bar B} \to X_s \gamma)^{\rm SM}_{E\gamma > E_0}
\nonumber \\
&+& B_{77} \frac{|\Delta C_7|^2}{|C_7^{\rm SM}|^2}
+ (2 B_{77}+B_{27}+B_{78})
Re\left[\frac{\Delta C_7}{C_7^{\rm SM}}\right]~,
\label{eftbsg}
\end{eqnarray}
where the numbers $B_{ij}=B_{ij} (\mu , \delta)$ are given
in \cite{kagan}.  For $\mu = m_b$, and $\delta =0.3$
($E_0 = (1-\delta)E^{\rm max}_\gamma =1.6$ GeV,
$E^{\rm max}_\gamma = m_b/2$) we have $B_{77}=0.361$,
$B_{27}=1.387$ and $B_{78}=0.08$.
We therefore have (with $C_7^{\rm SM}(m_W)=-0.22$ at NLO):
\begin{eqnarray}
10^4  BR (B \to X_s \gamma)^{\rm th}_{E\gamma > 1.6 \rm GeV} &=&
10^4 BR (B \to X_s \gamma)^{\rm SM}_{E\gamma > 1.6 \rm GeV}
\nonumber \\
&+& 7.4 |\Delta C_7 (m_W)|^2 - 9.9
Re\left[\Delta C_7 (m_W) \right]~.
\label{bsgformula}
\end{eqnarray}

The contribution from the dipole operator $Q_{tB}$
($=Q_{uB}^{33}$ as defined in \cite{grz10}) enters through
the electromagnetic dipole vertex in the penguin diagram.
Even before the top quark was observed a calculation
was done in Ref.~\cite{hewett}.  Their result was that
at the scale $\mu = m_W$ (with $x=m_t^2/m^2_W$):
\begin{eqnarray}
\Delta C_7 (m_W) &=& \kappa G_1 + i{\widetilde{\kappa}} G_2
  \nonumber \\
&=& \sqrt{2}
\frac{m_t}{m_W} \frac{v^2}{\Lambda^2} \frac{c_w}{s_w}
\left[ C_{tB} \frac{G_1+G_2}{2} + C^*_{tB} \frac{G_1-G_2}{2}
  \right] \label{hewettc7}~, \\
\frac{G_1+G_2}{2} &=& -\frac{1}{4} +\frac{1}{4}\frac{1}{x-1}
+ \frac{1}{8} \frac{x^2+x}{(x-1)^2} -\frac{1}{4}
\frac{3x-2}{(x-1)^3} \ln{x} ~,\nonumber \\
\frac{G_1-G_2}{2} &=& \frac{1}{8} - \frac{3}{8} \frac{1}{x-1}+
\frac{1}{4}\frac{1}{(x-1)^2} + \frac{1}{4}
\frac{x-2}{(x-1)^3} \ln{x} ~.\nonumber
\end{eqnarray}
This result is to be compared with the more recent one in
\cite{crivellin15} (scale $\mu_W = m_W$):
\begin{eqnarray}
\Delta C_7 (m_W) &=& \sqrt{2}
\frac{m_t}{m_W} \frac{v^2}{\Lambda^2} \frac{c_w}{s_w}
\left[ C_{tB} E^{uB}_7 + C^*_{tB} F^{uB}_7
  \right] \label{crivellinc7}~, \\
E^{uB}_7 &=& -\frac{1}{16} \frac{(x+1)^2}{(x-1)^2}
-\frac{1}{8} \frac{x-3}{(x-1)^3} x^2 \ln{x} \nonumber~, \\
F^{uB}_7 &=& -\frac{1}{8} \nonumber~.
\end{eqnarray}
We point out that these results from \cite{crivellin15} are the ones
that have been used in the literature in the last six years, as in
\cite{bissmannbsg}.  The expressions (\ref{hewettc7}) and
(\ref{crivellinc7}) are noticeably different both in their analytical
expressions and numerical values. For instance, with
$m_t=174$GeV we have
\begin{eqnarray}
  \frac{G_1+G_2}{2} &=& -0.030 \; ,\;\;\; \frac{G_1-G_2}{2} =0.062
  \nonumber \\
E^{uB}_7 &=& -0.183 \; ,\;\;\;\;\;\;\;\;\; F^{uB}_7 = -\frac{1}{8} \; ,
  \nonumber
\end{eqnarray}
which means that the results from \cite{crivellin15} make the
branching fraction much more sensitive to $C_{tB}$ (or $\kappa$) than
those from \cite{hewett}.  It is beyond the scope of this work to revise
these one-loop calculations in this case, though we may address
this issue in some future study.  For the present analysis, however,
we will use both results.
In terms of $\kappa$ and $\widetilde{\kappa}$, for $m_t=174$GeV we
have:
\begin{eqnarray}
  \cite{hewett}
\Delta C_7 (m_W) &=& 0.032 \kappa -i0.092 {\widetilde{\kappa}}~,
\label{c7khewett} \\
\cite{crivellin15}
\Delta C_7 (m_W) &=& -0.416 \kappa -i0.166 {\widetilde{\kappa}}~.
\label{c7kcrivellin}  
\end{eqnarray}
Inserting the recent SM value of $a^{\rm SM}=3.36$ and either
eq.~(\ref{c7khewett}) or eq.~(\ref{c7kcrivellin})
into eq.~(\ref{bsgformula}), we obtain: 
\begin{equation}
  \label{bsginkappas}
10^4 BR (B \to X_s \gamma)^{\rm theo} = 3.36 \pm 0.23
+ BR_{\kappa , \widetilde{\kappa}}~,  
\end{equation}
with
\begin{eqnarray}
\cite{hewett}BR_{\kappa , \widetilde{\kappa}} &=&
 -0.32 \kappa + 0.01 \kappa^2 + 0.06 {\widetilde{\kappa}}^2
\label{br.hewett}~,  \\
\cite{crivellin15}BR_{\kappa , \widetilde{\kappa}} &=&
4.1 \kappa + 1.3 \kappa^2 + 0.2 {\widetilde{\kappa}}^2
\label{br.crivel}~.
\end{eqnarray}

Before writing down the formulas for the allowed parameter
region, we would like to point out that as we were aware of the
recent study of constraints on $C_{tB}$ from $t{\bar t}\gamma$
and $B \to X_s \gamma$ in \cite{bissmannbsg}, we made a comparison
of our results with theirs.  We first notice that they use the
NP contribution $\Delta C_7$ by \cite{crivellin15}
(they define a coefficient ${\tilde C}_{uB}$ that is
equal to $\kappa / 5.62$),  which is written
in our eq.~(\ref{c7kcrivellin}).  Then, they plot the dependence
of ${\rm BR} (B \to X_s \gamma)$ on ${\tilde C}_{uB}$ which
has the shape of a parabola.
We have compared the ${\rm BR} (B \to X_s \gamma)$ we obtain
by using eq.~(\ref{bsgformula}) with their plot and we have found
good agreement.

The difference between the SM and the experimental
values is $3.36-3.32 =0.04$ with an uncertainty given
by $\sqrt{0.15^2 + 0.23^2} = 0.28$.  We can then set, 
at the $1\sigma$ C.L.:
\begin{eqnarray}
-0.04-0.28 \leq BR_{\kappa , \widetilde{\kappa}} \leq -0.04+0.28
\; .\label{bsgregion}
\end{eqnarray}
Thus, by setting ${\widetilde{\kappa}}=0$ we get from  $B \to X_s \gamma$:
\begin{eqnarray}
  \cite{hewett}:&\qquad
  -0.75 \leq {\kappa} \leq 1.0 \label{kappalimits.1}~,\\
  \cite{crivellin15}:&\qquad
  -0.08 \leq {\kappa} \leq 0.06
  \label{kappalimits.2}~,
\end{eqnarray}
which are the individual-coupling limits on $\kappa$ from the
branching ratio $BR ({\bar B} \to X_s \gamma)$.

The above limits for $\kappa$ are set at the scale $m_W$.
We can now obtain the corresponding limits at the scale $m_t$.
To do so, we just use equation~(\ref{rgecub}):
$\kappa (m_t) = 1.06 \kappa (m_W)$.
Let us rewrite equations (\ref{br.hewett}) and (\ref{br.crivel})
at the scale $m_t$ (define $k_t \equiv \kappa (m_t)$):
\begin{eqnarray}
\cite{hewett}BR_{\kappa_t , \widetilde{\kappa}_t} &=&
 -0.302 \kappa_t + 0.009 \kappa_t^2 + 0.053 {\widetilde{\kappa}_t}^2
\label{br.hewett.top}~,  \\
\cite{crivellin15}BR_{\kappa_t , \widetilde{\kappa}_t} &=&
3.87 \kappa_t + 1.16 \kappa_t^2 + 0.178 {\widetilde{\kappa}_t}^2
\label{br.crivel.top}~.
\end{eqnarray}
Thus, by setting ${\widetilde{\kappa}_t}=0$ we get:
\begin{eqnarray}
  \cite{hewett}:&\qquad
  -0.78 \leq {\kappa_t} \leq 1.10 \label{kappatlimits.1}~,\\
  \cite{crivellin15}:&\qquad
  -0.085 \leq {\kappa_t} \leq 0.064
  \label{kappatlimits.2}~,
\end{eqnarray}

\subsection{The $CP$ asymmetry ${\cal A}^{}_{Xs\gamma}$.}
\label{sec:asym}

The expression for the $B \to X_s \gamma$ $CP$ asymmetry can be
written as \cite{benzke}:
\begin{eqnarray}
    {\cal A}^{}_{Xs\gamma} &=& \left[\; \left( \frac{40}{81}-
      \frac{40}{9} \frac{\Lambda_c}{m_b} \right) \alpha_s + \pi
\frac{\Lambda^c_{17}}{m_b} \; \right] \; {\rm Im}\frac{C_1}{C_7}
    \nonumber \\
  &-& \left( \frac{4}{9}\alpha_s + \frac{4}{3} \pi^2 \alpha_s
     \frac{\Lambda_{78}}{m_b} \right) \; {\rm Im}\frac{C_8}{C_7}
     \nonumber \\
  &-& \left( \alpha_s \frac{40}{9}\frac{\Lambda_c}{m_b} + \pi
     \frac{\Lambda^u_{17} -\Lambda^c_{17}}{m_b} \right) \;
          {\rm Im}\left(\epsilon_s \frac{C_1}{C_7} \right) \, ,
     \label{benzkeasym}    
\end{eqnarray}
where the coefficients $C_j$ and $\alpha_s(\mu)=0.3$
at the scale $\mu = 2$ GeV are given by \cite{benzke}:
\begin{eqnarray}
C_1 (\mu) = 1.204 \; ,\;\; C_8 (\mu) = -0.175 \; ,\;\;
C_7 (\mu) = -0.381 + 0.55 \Delta C^{}_7(m_W) \; ,
\label{asycoeff} 
\end{eqnarray}
with $\Delta C^{}_7$ given in (\ref{c7khewett}),
(\ref{c7kcrivellin}), and
where $0.55=\eta^{16/23}$ is the factor for the running
from the $m_W$ scale down to $\mu = 2$GeV \cite{kagan}.
In addition $\Lambda_c \simeq 0.38$ GeV, $m_b=4.6$;
and $\epsilon_s$, by using the Wolfenstein parameters
in \cite{benzke}, is given by:
\begin{eqnarray}
\epsilon_s = \frac{V_{ub} V^*_{us}}{V_{tb} V^*_{td}} =
10^{-2} \left( -0.8 + i 1.8\right)\; . \label{epsilons}
\end{eqnarray}
The $\Lambda$ parameters are not known and in \cite{benzke} they were
given some limits that have recently been revised.  The parameter with
the greatest uncertainty is $\Lambda^u_{17}$:
$-660 < \Lambda^u_{17} < 660$ MeV \cite{ayesh}.  The other parameters
have an allowed range smaller by two orders of magnitude:
$-7 < \Lambda^c_{17} < 10$ MeV and $17 < \Lambda_{78} < 190$ MeV
\cite{ayesh}.

We can now write ${\cal A}^{}_{Xs\gamma}$ in (\ref{benzkeasym})
with the numerical values in \cite{benzke}:
\begin{eqnarray}
  10^2 {\cal A}^{}_{Xs\gamma} &=&
\left( 6.91\times 10^{-2} + 3.8 x_c + 0.7 x_{78} \right)
\; {\rm Im} \left[ \frac{1}{C_7} \right] \nonumber \\
&+& \left( 0.133 + 3.78 x_{uc} \right)
\; {\rm Im} \left[ \frac{-\epsilon_s}{C_7} \right]
\; ,\label{cpasymin} \\
x_u &=& \frac{\Lambda^u_{17}}{m_b}\;, \;\;
x_c = \frac{\Lambda^c_{17}}{m_b}\;, \;\;
x_{uc} = x_u - x_c \;, \;\;
x_{78} = \frac{\Lambda_{78}}{m_b}
\nonumber
\end{eqnarray}
As we have two different values for $\Delta C_7$, will
obtain two different evaluations of the asymmetry:
\begin{equation}
  \label{asyminkappas.1}
  \begin{aligned}
\cite{hewett}  10^2 {\cal A}^{}_{Xs\gamma} &= |C_7|^{-2}
\left( 0.091 + 0.355 {\widetilde{\kappa}} - 0.004 \kappa
\right)  \; ,\\
|C_7|^2 &= 10^{-2} \left[(3.81 -0.176 \kappa)^2 +
 0.256 {\widetilde{\kappa}}^2 \right] \; ,
  \end{aligned}
\end{equation}
and
\begin{equation}
  \label{asyminkappas.2}
  \begin{aligned}
\cite{crivellin15}  10^2 {\cal A}^{}_{Xs\gamma} &= |C_7|^{-2}
\left( 0.091 + 0.64 {\widetilde{\kappa}} + 0.055 \kappa
\right)  \; , \\
|C_7|^2 &= 10^{-2} \left[(3.81 +2.29 \kappa)^2 +
  0.834 {\widetilde{\kappa}}^2 \right] \; .
  \end{aligned}
\end{equation}
In both cases, (\ref{asyminkappas.1}) and (\ref{asyminkappas.2}), by
setting $\kappa = {\widetilde{\kappa}}=0$ we obtain
$10^2 {\cal A}^{\rm SM}_{Xs\gamma} = 0.627$.  There is an estimated
$\pm 2.6$ theoretical uncertainty in $10^2 {\cal A}^{}_{Xs\gamma}$
that comes from $\Lambda^u_{17}$.
The experimental averaged value for the asymmetry can be found
in \cite{PDG}:
$10^2 {\cal A}^{\rm exp}_{Xs\gamma} = 1.5 \pm 1.1$.
After adding the two uncertainties in quadrature we obtain a
global, ($1\sigma$) 68\% C.L.,  $\pm 2.8$ uncertainty and we can write
the inequality:
\begin{eqnarray}
-2.8 +1.5 \leq 10^2 {\cal A}^{}_{Xs\gamma}
\leq 2.8 +1.5 \label{asymregion}
\end{eqnarray}

In order to find the individual limits on $\widetilde{\kappa}$ from
both values of $C_7$ we set $\kappa =0$ and
approximate $|C_7|^2 \simeq 0.381^2$ in (\ref{asyminkappas.1}),
(\ref{asyminkappas.2}) to obtain: 
\begin{equation}
  \begin{aligned}
  \cite{hewett}:&\qquad
  -0.79 \leq  {\widetilde{\kappa}}  \leq 1.5 \\
\cite{crivellin15}:&\qquad
-0.44 \leq {\widetilde{\kappa}}  \leq 0.83
  \end{aligned}
\end{equation}

As before, let us now remember that so far the coupling $\kappa$
has been set at the scale $m_W$.  Replacing
$\kappa$ by $\kappa_t / 1.06$ in eqs.\ 
(\ref{asyminkappas.1})-(\ref{asyminkappas.2}), we obtain:

\begin{equation}
  \label{asyminkappas.1top}
 \begin{aligned}
\cite{hewett}  10^2 {\cal A}^{}_{Xs\gamma} &= |C_7|^{-2}
\left( 0.091 + 0.335 {\widetilde{\kappa}_t} - 0.0038 \kappa_t
\right)  \; ,\\
|C_7|^2 &= 10^{-2} \left[(3.81 -0.166 \kappa_t)^2 +
 0.228 {\widetilde{\kappa}_t}^2 \right] \; ,
  \end{aligned}
\end{equation}
and
\begin{equation}
  \label{asyminkappas.2top}
  \begin{aligned}
\cite{crivellin15}  10^2 {\cal A}^{}_{Xs\gamma} &= |C_7|^{-2}
\left( 0.091 + 0.604 {\widetilde{\kappa}_t} + 0.052 \kappa_t
\right)  \; , \\
|C_7|^2 &= 10^{-2} \left[(3.81 +2.160 \kappa_t)^2 +
  0.742 {\widetilde{\kappa}_t}^2 \right] \; .
  \end{aligned}
\end{equation}
From these equations, by setting $\kappa_t=0$ we obtain the bounds, 
\begin{eqnarray}
  \cite{hewett}:&\qquad
  -0.84 \leq  {\widetilde{\kappa}_t}  \leq 1.67 \label{asymzz.1}\\
\cite{crivellin15}:&\qquad
-0.47 \leq {\widetilde{\kappa}_t}  \leq 0.93\label{asymzz.2}
\end{eqnarray}

We point out that in Ref.~\cite{cirigliano16} limits on
$\widetilde{\kappa}$ that are three orders of magnitude
stronger have been reported:
\begin{equation}
|{\widetilde{\kappa}}| \leq 1.4 \times 10^{-3}~,
  \label{edmlimits}
\end{equation}
based on nuclei and electron electric-dipole moment measurements.

\section{Single-coupling bounds for Wilson coefficients}
\label{sec:app.bound}

In this appendix we write the most relevant results from tables
\ref{tab:bound.cpq}--\ref{tab:bound.cubr.2} for the couplings $\cmpq$,
$\cub$ associated with the basis operators $\mathcal{O}^{33}_{uB}$,
$\mathcal{O}_{\vphi q}^{(-)33}$ in terms of the couplings $\ccub$,
$\ccmpq$ associated with $Q^{33}_{uB}$, $Q_{\vphi q}^{(-)33}$.  Doing so
is useful to compare with some results in the literature. From those
tables and (\ref{eq:coupl2}) we obtain the limits on Wilson
coefficients given below at the LHeC energy. Results for the FCC-he
energy are completely analogous, as seen from the tables in sections
\ref{sec:cmpq.results}, \ref{sec:cub.results.1}. 

The limits on $\ccmpq$ are,
\begin{center}
  \begin{tabular}{ccc}\cline{2-3}
    &\multicolumn{2}{c}{$\ccmpq$, LHeC, $PhP_{III}$}\\\cline{2-3}
$\varepsilon_\mathrm{exp}$&68\% C.L.&95\% C.L.\\\hline
12\%&-0.58,0.64&-1.10,1.37\\
15\%&-0.71,0.81&-1.37,1.81\\
18\%&-0.86,0.99&-1.62,2.31\\\hline
  \end{tabular}
\end{center}

The limits on $\cub$ are,
\begin{center}
  \begin{tabular}{ccc|cc}\cline{2-5}
    &\multicolumn{2}{c}{$\ccubr$, LHeC, $PhP_{I}$}
    &\multicolumn{2}{c}{$\ccubi$, LHeC, $PhP_{I}$}\\\cline{2-5}
$\varepsilon_\mathrm{exp}$ & 68\% C.L.  & 95\% C.L. & 68\% C.L. & 95\% C.L.\\\hline            
12\%&-0.24,0.29&-0.45,0.65&$\pm$0.89&$\pm$1.24\\    
15\%&-0.30,0.37&-0.54,1.00&$\pm$0.94&$\pm$1.36\\    
18\%&-0.35,0.46&-0.65,3.54&$\pm$1.06&$\pm$1.48\\\hline    
\end{tabular}
\end{center}
These values are to be compared, e.g., with table \ref{tab:global} and
with \cite{bou13b}.

\end{document}